\documentclass[12pt]{article}
\usepackage{graphicx}
 \usepackage{longtable}
\usepackage{multirow}
\relax
\usepackage{amssymb}

\newcommand{\bo}{{\bar o}}

\def\bo{{\raise.15ex\hbox{\large$\Box$}}}               

\def\face{{\raise.2ex\hbox{$\displaystyle \bigodot$}\mskip-2.2mu \llap {$\ddot
        \smile$}}}                                      

\def\Zbf{{\bf Z}}

\def\leftrightarrowfill{$\mathsurround=0pt \mathord\leftarrow \mkern-6mu
        \cleaders\hbox{$\mkern-2mu \mathord- \mkern-2mu$}\hfill
        \mkern-6mu \mathord\rightarrow$}       
\def\dvec#1{\vbox{\ialign{##\crcr
        \leftrightarrowfill\crcr\noalign{\kern-1pt\nointerlineskip}
        $\hfil\displaystyle{#1}\hfil$\crcr}}}           

\def\beq{\begin{equation}}
\def\eeq{\end{equation}}

\def\beqx{\begin{displaymath}}
\def\eeqx{\end{displaymath}}

\def\beql{\begin{eqnarray}}
\def\eeql{\end{eqnarray}}

\newcommand{\Tr}{{\rm Tr}}

\newcommand{\bea}{\begin{eqnarray}}
\newcommand{\eea}{\end{eqnarray}}

\def\[{\left [}
\def\]{\right ]}
\def\({\left (}
\def\){\right )}

\def\+{\oplus}

\begin{document}

\hbox{\hskip 12cm NIKHEF/2010-30  \hfil}
\hbox{\hskip 12cm IFF-FM-2010/02  \hfil}
\hbox{\hskip 12cm September 2010  \hfil}

\vskip .5in

\begin{center}
{\Huge \bf Asymmetric Gepner Models} \\  \vskip .2truecm {\Large  \bf II. Heterotic Weight Lifting}

\vspace*{.4in}
{ B. Gato-Rivera}$^{a,b,}\footnote{Also known as B. Gato}$
{and A.N. Schellekens}$^{a,b,c}$
\\
\vskip .2in

${ }^a$ {\em NIKHEF Theory Group, Kruislaan 409, \\
1098 SJ Amsterdam, The Netherlands} \\

\vskip .2in

${ }^b$ {\em Instituto de F\'\i sica Fundamental, CSIC, \\
Serrano 123, Madrid 28006, Spain} \\

\vskip .2in

${ }^c$ {\em IMAPP, Radboud Universiteit,  Nijmegen, The Netherlands}

\end{center}

\vspace*{0.3in}
{\small
A systematic study of ``lifted" Gepner models is presented. Lifted Gepner models are obtained from standard
Gepner models by replacing one of the N=2 building blocks and the $E_8$ factor by a modular isomorphic $N=0$ model 
on the bosonic side of the heterotic string. The main result is that after this change three family models occur abundantly, in sharp
contrast to ordinary Gepner models. In particular, more than 250 new and unrelated moduli spaces of three family models
are identified.  We discuss the occurrence of fractionally charged particles in these spectra. 
}


\vskip 1in

\noindent
\newpage

\section{Introduction}

In a previous paper  \cite{GatoRivera:2010gv} an analysis   of spectra
of simple current invariants of Gepner models \cite{Gepner:1987qi} was presented, completing work that was initiated in 1989 in 
\cite{Schellekens:1989wx}. The motivation for this new,  extended analysis were the statistical results on the number of families in 
orientifold models. The observed number of families, three, turned out to occur two to three orders of magnitude less
frequently \cite{Dijkstra:2004cc,Gmeiner:2005vz} than the number one, two or four. We wanted to verify if the same suppression 
also occurred in heterotic models.

This return to the heterotic string forced us to face the problem which led one of us to essentially abandon the subject after 1989: the
appearance of particles with fractional electric charge in the spectrum, an undesirable feature first observed in a class of $(2,2)$ 
geometric compactifications \cite{Wen:1985qj,Athanasiu:1988uj}, and then generally proved to occur for all $(0,2)$ compactifications 
with the standard $SU(3)\times SU(2)\times U(1)$ embedding \cite{Schellekens:1989qb} based on exact CFT constructions. 

The results of  \cite{GatoRivera:2010gv} can be summarized as follows. No substantial progress was made on the issue of family numbers, despite a 
huge enlargement of the class considered:  as before, three families only occur for the exceptional model usually denoted as 
$(1,16^*,16^*,16^*)$ \cite{Gepner:1987hi}, and for any other exceptional or non-exceptional combination the number of families is always 
even (see also \cite{Fuchs:1989yv}). Hence it appeared that also in the heterotic case, three families is a rare feature. 
On fractional charges a somewhat 
more optimistic conclusion was reached,
namely that they are fairly often vector-like. Within the context of RCFT (rational conformal field theory) that is about the best one can
reasonably hope for.

However, this analysis led to a new idea, presented in \cite{GatoRivera:2009yt}, and called ``Heterotic Weight Lifting". 
Essentially this enables us to go one step further away from the lamppost of heterotic $(2,2)$ models, and get closer towards genuine
$(0,2)$ models. In  \cite{Schellekens:1989wx} this was done by not imposing the equivalent of space-time supersymmetry in
the bosonic sector of the heterotic string; in  \cite{GatoRivera:2010gv} the world-sheet supersymmetry constraints between CFT
building blocks were relaxed in the bosonic sector, and furthermore the $SO(10)$ group was broken (see also \cite{Blumenhagen:1995tt}). 
Heterotic weight lifting
allows us to replace a minimal N=2 model plus the $E_8$ gauge group in the bosonic sector by a new CFT that has exactly
the same modular $S$ and $T$ matrices. This implies that modular invariance is manifestly preserved. However, the spectrum
is drastically changed: the conformal weight of all ground states is shifted by integers, and the dimension
of these ground states is changed as well. Because the weight of most ground states is moved up, we called this procedure
``weight lifting". However, there are also some ground states that move down. Therefore, in order to be able to compute
{\it massless} spectra of lifted Gepner models it is essential to be able to compute the {\it exact} spectra of standard Gepner models. 
This is indeed possible in the constructions considered here.
In \cite{GatoRivera:2009yt} a few examples of lifted Gepner models were presented, which indicated in particular that
the chances for getting three families looked better. Here we want to present a complete analysis of all 435 cases of heterotic
weight lifting one obtains by applying the list of lifts shown in \cite{GatoRivera:2009yt} to the 168 Gepner models.
We will focus on the same two issues considered in  \cite{GatoRivera:2010gv}: the number of families and the presence of fractional charges. 
On the first issue our conclusion is much more positive than the one in our previous paper, whereas on the second issue the
conclusion remains essentially unchanged.

This paper is organized as follows. In the next section we explain our motivations. Readers who are already sufficiently motivated may
skip this section, but  our perspective  of string phenomenology is rather different from that of others working in this area, and so we believe 
some explanation may be useful. In section three we summarize the heterotic weight lifting procedure. In section four we present our
results on charge quantization and in section five the results on the number of families. Section six contains results on distributions
of vector-like particles. In section seven  we investigate whether matter in this class of models  generically couples to the
extra non-abelian groups provided by the heterotic weight lifting procedure, instead of the familiar $E_8$ factor. Finally, in section
eight we formulate our conclusions.

\section{Motivation}

This paper is about RCFT, and by their nature such constructions are more suited for studying the structure of the standard model
(gauge groups and representations) than for parameters and moduli. 
Obviously the first question to ask regarding the standard model structure is: can one build exact
string theories that have a gauge group $SU(3)\times SU(2)\times U(1)$, three families, and no fractionally
charged exotics? It has been amply demonstrated that the answer to this question is positive, both in the case of orientifolds and
heterotic strings (see our previous paper \cite{GatoRivera:2010gv} for references).
The common attitude in string phenomenology seems to be that this is all we need to know about
the standard model structure, and that we should move on
to issues like parameter values and moduli stabilization. 
Clearly that {\it is} indeed important. But there are other questions regarding the standard model structure we should
not ignore.

A couple of decades ago, it was hoped that string theory would determine the standard model uniquely. 
Already very soon that turned
out to be 
unrealistic, and it  was
replaced by the hope that some mysterious missing selection principle would uniquely select the standard model. Furthermore it was
hoped that 
while waiting for the discovery of that principle, one could at least try to make a complete enumeration of all candidate vacua, and
find the one that agrees with all available data, in particular the standard model
and all of its parameters. Having identified it, a plethora of predictions would be available, and the 
correctness of the theory would be proved unambiguously by testing those predictions. 

This point of view -- although rarely expressed so explicitly -- could still be maintained until the beginning of this century, but it seems
now unlikely that we will be in that situation in the foreseeable future, because
the amount of  string vacua is likely to vastly exceed the currently available amount of experimental data
(since the latter increases monotonically, this balance may tip in the future).
One may still hope
that there is something fundamentally wrong with our understanding of moduli stabilization, supersymmetry or the cosmological
constant, and  that we eventually will end up with a much smaller set. One may also hope that the large numbers are somehow 
irrelevant to the problem
at hand ({\it i.e.} getting the standard model from string theory), and that they can be cleanly factored out from the problem, 
contributing only to the tuning of the cosmological constant. Even if that were true, we are still facing the problem that the 
methods we use to explore the
string theory landscape are primitive and limited. On the exact CFT side they are mostly limited to non-interacting CFT's 
(free bosons, free fermions and orbifolds), a very special case, whereas geometric approaches to heterotic strings are 
constrained to the neighbourhood of $(2,2)$ models. 
In other words, at present the chances of finding ``the standard model"
in string theory and make exact predictions based on it seem very small. A possibly achievable goal is to prove that the landscape
is dense enough in all relevant directions to contain the standard model (and anything discovered beyond it at a given time), and perhaps this
is all that can be done in the foreseeable future.

But one may still hope to be able to make generic predictions for {\it classes} of models. 
Possible examples are extra $U(1)$'s, rank-2 tensors
that are ubiquitous in orientifold models, mirror fermions or particles 
with fractional electric charge. Such predictions based on generic features of certain classes will never be absolutely falsifiable, but
if such particles are found they will nevertheless boost the confidence in the theory. Conversely, if it can be shown that certain particles are 
generically present in string theory, but are never found, one should start having serious doubts about either string theory in general, or
specific classes of string theories.
However, before making such predictions, it would be prudent to check if similar ones would have worked in the past. For example,
had there not already been strong limits on fractionally charged particles a few decades ago, the early results of heterotic strings 
might have led to predict their existence. This prediction, which is actually a postdiction, would have failed. 
One should be highly skeptical of {\it pre}dictions if the {\it post}dictions are wrong. In other
words, one should not only demand that the standard model is among the string vacua in a class, but also that its observed features  are
not excessively rare in that class. If indeed fractional charges are abundant in most heterotic string vacua that otherwise produce the
standard model correctly, then this should be viewed as a serious problem.

Most of string phenomenology is focusing on getting the observed features out while ignoring
the neighbouring landscape. We would argue that scanning the landscape in the neighbourhood of the standard model is 
essential in order to have any confidence in the correctness of a certain approach.  If certain features systematically come out wrong,
and can be gotten right only in very special and rare conditions, then we must assume that there is something missing in our understanding. 
Such failures may be due to a number of causes. Perhaps one is looking at the wrong kind of string theory (orientifolds, heterotic strings, F-theory)
or one is seeing artifacts of a computation method ({\it e.g.} free fields or RCFT), or one uses a too severe approximation to a true string vacuum
(for example unbroken supersymmetry or unstabilized moduli). It is also possible that certain rare features are {\it needed} for the existence of 
(intelligent) life, and hence it is not surprising that we observe them. If all this fails one can always claim that apparently we ended  up in a 
universe that has some  exceptionally rare features for reasons that we cannot figure out. In our opinion this is
the least attractive option. 
In any case it is important to confront these problems, rather than
ignoring them by focusing solely on string vacua that fit the observations. 

Anthropic arguments, in the sense defined above, are unavoidable and perfectly acceptable
in the context of a landscape   \cite{Linde:1986fd,Susskind:2003kw,DutchText,Schellekens:2008kg}, if indeed it can be 
convincingly demonstrated that a certain feature is or is not important for
the existence of life. 
But usually such arguments are just too complicated to be used reliably. What one
should do instead, before worrying about a rare feature, is to try to be as certain as possible that  at least anthropic arguments are not
likely to make a difference. In \cite{GatoRivera:2010gv} we discussed this issue for the number of families and the presence of 
fractionally charged particles: although it cannot be demonstrated rigorously, it is hard to believe that the existence of life requires 
three families (as opposed to two or four) or is inhibited by the mere existence of very rare, and heavy, fractionally charged 
particles (on the other hand, if they are either light
or abundant the argument is far too complicated to arrive at decisive general conclusions).
We would be perfectly happy to be
convinced otherwise, but we take as our working hypothesis that anthropic arguments do not play a r\^ole here. 

Any study of distributions of standard model properties is plagued by bias and measure problems (see for example \cite{Dienes:2008rm}). 
Some features of distributions
may be artifacts of the
class considered or the method used. Furthermore distributions can be affected by factors beyond current knowledge, such as
cosmological selection or moduli stabilization. Bias effects can partly be avoided by investigating the same issues using different 
methods or in different regions of the landscape. 
For example, the paucity of three family models in orientifold constructions was observed both in Gepner orientifolds and in 
$\Zbf_2\times \Zbf_2$ orientifolds; some questions raised in the present paper have also been examined in the context of 
free fermion models; and the present paper itself examines if earlier results on Gepner models hold also in a 
different class of heterotic models. Any investigation of distributions makes the implicit assumption that 
unknown factors are smoothly behaved in the relevant region of parameter space. For example, there is no reason to believe that
cosmological selection or moduli stabilization favors three families, so it is difficult to believe that such 
effects will lift the  large dip in orientifold distributions.  As long as no such argument is found, the 
problem should be taken seriously, and not simply dismissed as a statistical fluke.

It is a matter of taste what one defines as ``rare" or ``unnatural". In the literature,
undesirable features at the $1\%$ level are either ignored or stressed, depending
on the point of view one wishes to express. Most people would perhaps not worry about
 a dip of a factor of hundred in the family distribution
of orientifold models, precisely at the value three. But on the other hand, most people would say that a class of models
predicts a certain property if $99\%$ of the set had that property. Likewise, convergence of gauge couplings with a precision at the
$1\%$ level is considered strong evidence by some, and ignored by others. The discrepancy of the GUT scale and the string scale
by two orders of magnitude, on the other hand, is considered small by adepts of the heterotic string, 
and taken to infinity in discussions on F-theory GUTs.

\subsection{Remarks on vector-like particles}

Essentially all string spectra that produce the known particles contain large numbers of
additional particles: singlets, mirror quarks and leptons, higher rank tensors (in orientifold models)  and in some cases even more exotic 
ones such as fractionally charged particles. It is reasonable to require that all additional particles should be 
vector-like\rlap,\footnote{With a slight abuse of terminology  (since we apply the
term also to bosons) we will call all particles that can get a mass without breaking $SU(3)\times SU(2)\times U(1)$ ``vector-like".} and that all
matter that is chiral with respect to $SU(3)\times SU(2) \times U(1)$ has already been observed.
String theory provides robust
chiral spectra modulo non-robust vector-like states. To identify potential string vacua that match our world, we have to be able to match
at least the robust part of the spectra. 

However, it has become standard practice in string phenomenology to go much further and also match the non-robust part of the spectrum,
as if we are already certain that the currently known spectrum (plus the Higgs and all superpartners) is all that exists. In other words, one
assumes the absence of any vector-like particles other than those mentioned.
This is a remarkable assumption given that, on the one hand, some vector-like particles are expected to exist, but 
on the other hand we do not (yet) have
experimental information about any of them.

Exact RCFT constructions are more seriously affected by the problem of massless vector-like 
particles because they cannot be continuously deformed in any practical way. However, usually one gets a large number of discrete
points in what appears to be the same moduli space. 
This results in a discrete scan over possible masses of vector-like particles, which manifests itself in  the form of discrete distributions
of the number of massless vector-like particles. This phenomenon has been observed both in orientifold models and heterotic models.
If one is lucky enough the set includes points where certain vector-like particles are completely absent. However, it is
unlikely that the problem is solved entirely within the context of RCFT. This is simply not the right tool for solving this problem.
 Several steps are needed to move from
the exact RCFT to a phenomenologically acceptable string vacuum, and masses for vector-like particles may be generated at any of these
steps. They may acquire masses from simply moving away from the special RCFT point into the full moduli space, from moduli
stabilization or from supersymmetry breaking.  What is known about this is mainly folklore, based on common beliefs and
sporadic examples (see 
{\it e.g.} \cite{Faraggi:1990af,Buchmuller:2005jr,Lebedev:2006kn,Lebedev:2008un}
 for discussions of decoupling of exotic states in various models).   

Let us assume, for simplicity, a scenario where the physics of moduli stabilization is cleanly separated from the physics of 
supersymmetry breaking, and where the latter scale is low, in the LHC range.
It is quite generally accepted that all vector-like particles will acquire a mass after supersymmetry breaking 
(as well as the breaking of any
other symmetries beyond $SU(3)\times SU(2) \times U(1)$). It is less clear what a generic string spectrum looks like when all moduli
have been stabilized but supersymmetry is unbroken. If generically such a spectrum contains massless 
vector-like states beyond the MSSM, then
this implies a generic prediction of the existence of such vector-like states with masses generated by 
supersymmetry breaking, in addition to
the squarks, the sleptons and the gauginos. If the SUSY scale is low, these particles should be observed at the LHC. 
In this scenario, it makes little sense to focus on the rare cases where such particles are absent from 
the exact spectrum, because the chance
that these cases are realized in our universe is very small.   
On the other hand,
if the generic situation is that vector-like particles acquire large masses in generic points in moduli space or in 
moduli-stabilized vacua, it is pointless to worry about
their presence in exact RCFT. Many other scenarios with a variety of intermediate scales can be envisaged, 
but the conclusion remains unchanged in all cases where the supersymmetry breaking scale is low. 
If supersymmetry breaking is the last safety net to catch such particles, one would not expect it to be empty, 
unless something else catches them at an earlier stage. 

Running of coupling constants does impose some limits on light vector-like particles, even if they cannot (yet) be observed directly. 
In  exact string spectra there are often so many that 
if we extrapolate the low energy gauge couplings they develop Landau poles below the Planck scale; conversely, starting at
the Planck scale they would run to far too small values in the infrared.  
Furthermore vector-like particles
might ruin standard gauge coupling convergence, if they are not in complete GUT multiplets\rlap.\footnote{Note however that even for fractionally
charged exotics it is possible to find multiplets that have no impact on gauge coupling unification. An example is a standard model family 
with the opposite sign for all $Y$-charges, which gives rise to third-integral charges. Just like an extra family, this
affects only the value of the unified coupling, but not  the convergence itself.}  The observation of low energy couplings roughly in agreement with the SUSY-GUT scenario
does indeed eliminate many candidate heterotic string spectra,
but none of these arguments completely forbids the existence of light vector-like particles, and
they might even have the virtue of helping to reduce the  notorious GUT-scale/string scale gap.

However, light ({\it e.g.} near the TeV-scale) fractionally charged particles are almost certainly a 
phenomenological disaster, since they would be stable. The upper limit on their
abundance on Earth is far less than $10^{-20}$ per nucleon \cite{Perl:2009zz}, and stable light 
particles would be produced far too copiously in the early
universe to be consistent with that limit \cite{Athanasiu:1988uj}. 
Hence if we allow fractionally charged particles as vector-like massless particles in exact RCFT, we are making the
implicit assumption that in the full theory they acquire essentially Planck scale, or at least GUT scale masses 
(the lower limit depends
on various assumptions about the cosmological evolution of the universe).   
Provided this is what
generically happens, vector-like fractional charges are acceptable in an exact RCFT spectrum. 
At least one can give them the
benefit of the doubt.

If, on the other hand, generically some vector-like states remain light, this produces an interesting dilemma for heterotic strings. First of all
one would be forced to look for examples with a rare feature, {\it i.e.} complete absence of fractional charges in the exact RCFT massless spectrum,
with no argument why we find ourselves in such a special universe. Secondly it would undermine the usual assumption
that the low-energy spectrum should be that of the MSSM, and nothing more. Other vector-like particles do not satisfy the same
strict limits as stable, fractionally charged ones, and without having such limits, the logical 
conclusion would be to {\it predict} their discovery at the LHC, rather than bend string theory out of shape to fit
non-existent data.

We conclude that it is important to find out if  exact heterotic string spectra without massless fractional charges are indeed rare, and to do
so in as many different constructions as are accessible. We will discuss our results for heterotic weight lifting in section 4.

\section{Heterotic Weight Lifting}

The starting point of the construction of heterotic strings using interacting rational conformal field theory is
a diagonal modular invariant partition function with identical building blocks in the bosonic and the fermionic sectors.
In the fermionic sector the choice of building blocks is tightly constraint by the requirement of world-sheet supersymmetry.
But on the bosonic side the only constraint is conformal invariance, and hence a much larger set of building blocks is
in principle available. However modular invariance makes it nearly impossible to combine distinct building blocks in the
left and right sectors.

Heterotic weight lifting \cite{GatoRivera:2009yt} is a special case of a presumably very large set of modifications of the heterotic
string, based on the idea of replacing some building blocks in the bosonic sector by isomorphic ones in the
sense of the modular group. The difficulty is then to find building blocks that are isomorphic, but not identical, and
have the same central charge.
Some examples are known, such as the affine algebras of the $E_8 \times E_8$ and $SO(32)$ heterotic strings or
the meromorphic $c=24$ CFTs \cite{Schellekens:1992db}. All of these are CFTs with just a single primary field.
But what we
need here are building blocks that are isomorphic to $N=2$ minimal models.  

It is an interesting question whether there exist CFTs with the same $S$ and $T$ matrices and central charge as a given
$N=2$ minimal model, without being identical to it. We do not know of any example. However, we
can remove the superfluous $E_8$ factor that is present in the bosonic sector of a $(2,2)$ heterotic string and
allow the central charge to differ by 8. There is a well-known example of such an isomorphism: $SO(N)_1 \times E_8$ 
versus $SO(N+16)_1$. 

In a nutshell, the method works as follows.  We start with the level-k minimal
model times $E_8$. The minimal model can be realized as a coset CFT:
$$ \frac{SU(2)_k \times O(2)}{U_{2(k+2)}} $$ 
The precise construction of this CFT involves ``field identification", which can be viewed as a formal extension
of the chiral algebra with a spin-0 current of order 2.
Now we  ``deconstruct" this coset construction
by formally removing the field identification, then we embed the denominator algebra 
$U_{2(k+2)}$ in $E_8$ instead of $SU(2)_k \times O(2)$, and then we re-establish the field identification as a standard
simple current extension. 

The resulting CFT has a chiral algebra $SU(2)_k \times O(2) \times X_7$, where $X_7$ is the remainder of $E_8$ after
dividing out $U_{2(k+2)}$. In other words, the tensor product $U_{2(k+2)} \times X_7$ can be chirally extended to $E_8$. 
The factor $X_7$ has central charge 7, and its modular transformation matrices $S$ and $T$ are the complex conjugates
of those of $U_{2(k+2)}$. In \cite{GatoRivera:2009yt} a list of 33 such combinations was given, which was not claimed to be complete.
We will call the isomorphic $N=0$ CFT the ``lift" of the corresponding $N=2$ CFT. 
For some values of $k$ no such lift was found, and for some values there are two. Meanwhile we have attempted to find
more examples, but so far without success. Nevertheless, we feel that the list published in \cite{GatoRivera:2009yt} is just
the ``low-hanging fruit". These examples were easy to get, because they can be realized in terms of tensor products of
affine Lie algebras. The presence of these affine Lie algebras is important in order to reproduce the large simple
current groups of the minimal models, but there may well be other ways of achieving that.

Another possibility considered in \cite{GatoRivera:2009yt} is to replace {\it two} minimal model factors and the $E_8$ factor 
by an isomorphic CFT. One example was found, but it will not be considered here.

A Gepner model is characterized by a set of values $(k_0,\ldots,k_M)$ for the levels of the minimal model factors. 
In principle each of the values can be lifted, provided a lift exists. We will denote a lifted Gepner model by
$(k_0, \ldots, \widehat{k_i},\ldots,k_M)$ if the $i^{\rm th}$ factor is lifted. If $k_i$ has two distinct lifts, we use 
the notation $\tilde k_i$ for the
second one. There are 168 standard Gepner models. Applying all 33 single lifts listed in \cite{GatoRivera:2009yt}, we end
up with 435 lifted Gepner models.

Note that we may combine lifts with any modular invariant partition function of the standard Gepner models, including 
exceptional ones. Here we will not consider the latter. Apart from that, we will consider exactly the same set of MIPFs as
in our previous paper \cite{GatoRivera:2010gv}, allowing in particular the breaking of $SO(10)$ to any of the eight subgroups
listed there. We also allow breaking of world-sheet and space-time supersymmetry in the bosonic
sector. Of course this is already the case as soon as we replace one of the minimal models by its lift. However, in addition to that 
we also allow the space-time supersymmetry current and the world-sheet supersymmetry alignment currents of the
fermionic sector to be mapped to any isomorphic current in the bosonic sector.

\section{Fractional Charges and Group Types}

In \cite{GatoRivera:2010gv} we presented a detailed discussion of $SO(10)$ breaking and the associated fractional
charges (by which we always mean electric charges of color singlet particles).
In the present work, $SO(10)$ breaking
is treated in exactly the same way, and therefore {\it a priori} the possibilities are exactly the same. We just summarize
the main points for convenience. 

The heterotic string revolutionized string theory in 1984 \cite{Gross:1984dd} partly because it became
clear almost immediately  \cite{Candelas:1985en} that chiral four-dimensional spectra could be obtained with a remarkable
resemblance  to the observed particle spectrum. It is not an exaggeration to say that by merely
imposing chirality and four uncompactified space-time dimensions one is almost inescapably led to spectra consisting
of a number of families of $(16)$'s of $SO(10)$ (or in the special cases discovered earliest \cite{Candelas:1985en}, $(27)$'s of $E_6$). 
This is especially easy to demonstrate by using the ``bosonic string map" \cite{LLS}.

%

Unfortunately, this does not work so beautifully anymore if one tries to break $SO(10)$ to $SU(3)\times SU(2) \times U(1)$. 
This cannot
be done with a field-theoretic Higgs mechanism, because the required Higgs representations cannot occur as massless states
for affine $SO(10)$ at level 1. Instead
one can do the breaking directly in string theory.  In our language, the breaking amounts to 
writing the theory in terms of  $SU(3)\times SU(2) \times U(1)_{30}$ affine Lie algebras and not allowing
the $SU(5)$ roots to appear in the massless spectrum.  Here $U(1)_{2N}$ denotes a free boson compactified on a circle such that
there are $N$ primary fields with conformal weights $q^2/{4N}$, for $q=-N+1,\ldots,N$. Applying the standard group theory breaking
$SU(5) \supset SU(3) \times SU(2) \times U(1)$ to affine $SU(5)$ level 1 yields $SU(3)_1\times SU(2)_1 \times U(1)_{30}$ (we will
omit the level indices on $SU(3)$ and $SU(2)$ henceforth).
However, the only way to eliminate the $SU(5)$ roots while respecting modular invariance
is by allowing some of the fractionally
charged representations to appear. They must appear in the full theory, although not necessarily in the massless sector.
In other contexts (orbifolds, Calabi-Yau constructions) the breaking  is often achieved by using background gauge fields on Wilson lines, but no matter
how it is done, as long as the result can be described in terms of $SU(3) \times SU(2) \times U(1)_{30}$ as above, this description
applies.

Rather than breaking $SO(10)$ we start from its sub-algebra 
$SU(3) \times SU(2) \times U(1)_{30} \times U(1)_{20}$, which we extend by a simple current of order 30 to $SO(10)$ in the fermionic sector. 
In the bosonic sector we allow
some powers of this current to be replaced by other currents of the same order and relative monodromies. This allows
in total eight different subgroups of $SO(10)$, listed in Table \ref{TableG}. Which of these groups can be realized for a given tensor
product depends on the values of the levels $k_i$ exactly as described in \cite{GatoRivera:2010gv}, and independent of the
lifting. Each of these groups can still be further extended by means of currents with components outside $SO(10)$.

In this setup the first three factors,  $SU(3) \times SU(2) \times U(1)_{30}$, are destined to become the standard model gauge group.
Only a limited set of representations of this group 
has conformal weight less than or equal to one, and can occur in the massless spectrum. These are
the standard quark and lepton multiplets Q, U, D, L, E and their conjugates, singlets and
the $SU(3)\times SU(2)\times
U(1)$ representations $(3,2,\pm \frac56)$. The latter have weight one, and occur as extended roots in $SU(5)$ GUT models 
(yielding the familiar $X$ and $Y$ bosons). These representations do not occur as matter particles. All other
representations that can occur in the massless spectrum violate the standard model charge integrality sum rule $\frac{t}{3}+\frac{s}{2}+Y \in \Zbf$ (where
$t$ is $SU(3)$ triality and $s$ is the $SU(2)$ spin, modulo 2) and hence lead to color singlets with fractional electric charge. In the
following we reserve the name ``exotics" to particles that violate the charge integrality sum rule. 

The $SU(3)\times SU(2) \times U(1)$ embedding we consider here is the standard one which would be obtained if
one breaks $SO(10)$ to $SU(5)$ and then to $SU(3)\times SU(2) \times U(1)_Y$, with $U(1)_Y$ completely embedded in 
$SU(5)$. It is
singled out by the requirement that the couplings converge in the canonical $SU(5)$ way. This embedding offers the best opportunity
to understand the two ``GUT miracles" (coupling convergence and family structure) in a natural way in heterotic strings, and perhaps even in string theory in general.
It follows from the structure of the CFT that 
in this context the only fractional electric charges that can occur are $\frac12$, $\frac13$ and $\frac16$. There are many
other ways of embedding the standard model in the heterotic string. Examples are
flipped $SU(5)$ models, which provide a different way to embed the standard model in $SO(10)$. 
Furthermore, in $(2,2)$ models  one may consider  general embeddings of $U(1)_Y$ in $E_6$, as was done for example  
in \cite{Wen:1985qj} and \cite{Athanasiu:1988uj}. In these papers other fractional charges, like $\frac15$, are mentioned. 
This must be due to a different choice for the  $U(1)$ embedding.  Furthermore  
one may consider affine Lie algebras $SU(3)$ and $SU(2)$
at several higher levels, and distribute them freely over the available $c=22$ internal CFT. In the rest of this paper we will often use the
term ``heterotic strings" to refer to the specific $SU(3) \times SU(2) \times U(1)_{30}$ class described above.

\subsection{Group Types}

If the algebra $SU(3) \times SU(2) \times U(1)_{30}$ is extended,
the set of allowed fractional charges can be reduced, but it can only be reduced to integers if the extension
includes $SU(5)$. For each group type, the allowed fractional charges are listed  in the last two columns of Table \ref{TableG}.
The first of these columns lists the fractional
charges one would expect on the basis of the gauge group (given the fact that by construction the quantization of the $U(1)$ factor is
in units of $\frac16$), and
the second column lists the true quantization in string theory, as explained in  \cite{GatoRivera:2010gv}.

Of these eight groups, the $SU(5)$ GUT and the $SO(10)$ GUT are unacceptable, since, as explained above, they come without Higgs bosons
to break the GUT gauge symmetry (we do not consider the possibility that such Higgses are generated dynamically as bound states
of other massless matter).  Higgs bosons to break the Pati-Salam group or the left-right symmetric groups to the standard model
can exist in the massless spectrum, and hence these options will be considered viable. The most attractive possibility is nr. 2, called
``SM, Q=1/2", because it has a gauge group closest to the standard model, and a charge quantum closest to 1. 
Within the $SO(10)$ sector of the theory it has a gauge group which is exactly $SU(3)\times SU(2)\times U(1)$, times
a superfluous extra $U(1)$. 
In some cases this extra $U(1)$ is B-L, but there are other possibilities.

\renewcommand{\arraystretch}{1.2}
\begin{table}[]
\begin{center}
\vskip .7truecm
\begin{tabular}{|c|c|c||c|c|}
\hline
\hline
Nr. & Name & Gauge group  & Grp. & CFT \\
\hline
0 &\small SM, Q=1/6&   {\footnotesize  $SU(3)\times SU(2) \times U(1) \times U(1)$} &  $ \frac16$  & $ \frac16$  \\
1 &\small SM, Q=1/3 &  {\footnotesize $SU(3)\times SU(2)\times U(1)\times U(1)$}  &  $ \frac16$ & $ \frac13$  \\
2 &\small SM, Q=1/2 &  {\footnotesize $SU(3)\times SU(2)\times U(1)\times U(1)$}  &  $ \frac16$  & $ \frac12$  \\
3 &\small LR, Q=1/6 & {\footnotesize $SU(3)\times SU(2)_L \times SU(2)_R\times U(1)$}  &  $\frac16$  & $ \frac16$  \\
4 &\small SU(5) GUT&  {\footnotesize  $SU(5) \times U(1)$}  &  1  & 1  \\
5 &\small LR, Q=1/3 &   {\footnotesize $SU(3)\times SU(2)_L \times SU(2)_R\times U(1)$}  &  $\frac16$  & $ \frac13$  \\
6 &\small Pati-Salam &  {\footnotesize  $SU(4)\times SU(2)_L \times SU(2)_R$}  &  $\frac12$  & $ \frac12$  \\
7 &\small SO(10) GUT & {\footnotesize $SO(10)$} &  1 & 1 \\
 \hline
\end{tabular}
\vskip .7truecm
\caption{\small List of all standard model extensions within $SO(10)$ and the resulting group theory and
CFT charge quantization (last two columns). We refer to these subgroups either by the label in column 1 or by the name in column 2, where
``LR" stands for left-right symmetric.}
\label{TableG}
\end{center}
\end{table}
 
\begin{figure}[t]
\begin{center}
\includegraphics[width=4cm]{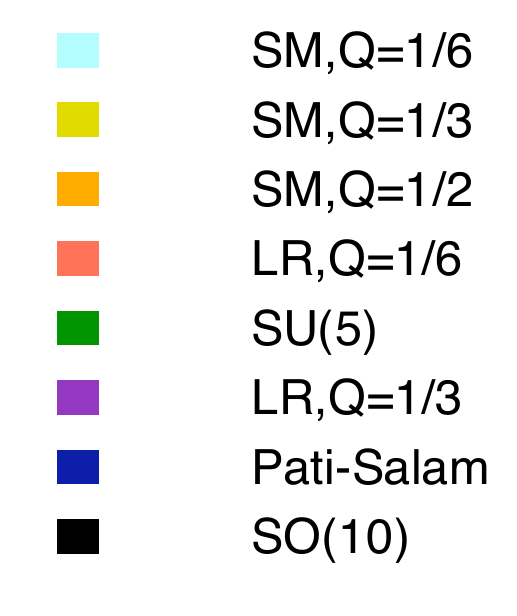}
\caption{Color codes for group types.}
\label{CC}
\end{center}
\end{figure}

In the figures we will distinguish these group types by means of the color codes\footnote{The color codes used
here are slightly different than those used in \cite{GatoRivera:2010gv} in order to make 
them more easily distinguishable in back and white printouts.} shown in Fig. \ref{CC}.
 In the text we will usually refer to them by the names in the second column of Table \ref{TableG}.

\subsection{Anomalies}

In heterotic strings of the type considered here, anomalies cancel by means of the standard Green-Schwarz mechanism involving
the $B_{\mu\nu}$ field. Note that  string theories obtained by heterotic weight lifting do not have a known geometric interpretation, and
cannot be derived in any known way by compactifying ten-dimensional heterotic strings. Hence the only way to
derive anomaly factorization here is by computing the ``elliptic genus" of the CFT, and using its modular properties, as 
explained in \cite{Schellekens:1986xh}.
 This computation applies in this case, because it is valid in complete generality for all $(0,2)$ CFT constructions (in fact
even for $(0,1)$ constructions). 

This result implies that the anomaly factorizes in the standard way in a factor linear in $F$ and $R$, times a factor 
$\Tr F^2 - \Tr R^2$, where $F$ is evaluated in the vector representation of an orthogonal group. Furthermore it follows
that $\Tr F^2$ receives contributions from every factor in the gauge group.  If the $E_8$ factor in the gauge group is unbroken, it
cannot produce a term of the form $\Tr F^2 \Tr F'$, where $F$ is an $E_8$ two-form and $F'$ a $U(1)$ two-form. 
This is because the only non-trivial representation of $E_8$ level 1 that can be massless is the
$(248)$, and since this exactly saturates the conformal weight, there is no room for a $U(1)$ charge. Hence there are no
massless representations charged under both $E_8$ and any $U(1)$. If $E_8$ cannot contribute to $\Tr F^2$, it follows that 
there cannot be a $\Tr F^2 - \Tr R^2$ factor at all, and hence the anomaly must be identically 0.
Since heterotic weight
lifting breaks the $E_8$ factor, it is no longer present in the gauge group. Therefore in general there can be a non-trivial
Green-Schwarz mechanism in these models.   This also implies that $U(1)$ factors can be anomalous.  This may happen in particular to the
standard model $U(1)$ factor $Y$ or the $U(1)$ corresponding to B-L. The corresponding $U(1)$ gauge boson will then become massive because
of the Stueckelberg mechanism, absorbing the $B_{\mu\nu}$ field.  
Note that  for this to happen, the $U(1)$ factor must really be anomalous. This is different from orientifold models: 
there even a non-anomalous $U(1)$ can acquire a mass from axion mixing. It follows that in a heterotic model with a certain number of 
standard model families, but no chiral exotics, $U(1)_Y$ must remain unbroken, because it is anomaly free. However, the extra
$U(1)$ that commutes with the standard model gauge group within $SO(10)$ can be anomalous. 
This is because we are not requiring this charge to coincide with B-L for all
quarks and leptons.  

In the absence of fractionally charged matter, anomaly cancellation would
force the spectrum into a certain number of standard model families. One question we will be interested in is to which extent
the existence of fractionally charged matter changes that.

\subsection{Group type distributions}

First we will examine how
often we find each group type if we simply select MIPFs at random. MIPF selection requires choosing a subgroup of the full set
of simple currents, plus a discrete torsion matrix defined on that subgroup \cite{Kreuzer:1993tf}. We have randomized this choice by
first choosing $N$ random elements of the full simple current group, with equal weight for each element. Then we compute the
subgroup  generated by these $N$ elements. Finally we either enumerate all
allowed discrete torsions defined on that subgroup if there are fewer than a hundred, or otherwise we take a random sample.  The amount
of time needed to compute spectra increases rapidly with $N$, and in practice values of $N$ larger than 4 are difficult to deal with. 
Our data are based on nearly saturated samples for $N=0,1$ and 2, but on incomplete random samples for N=3 and 4. Obviously the
precise results depend on various details of the sampling method, and in particular on the number of MIPF samples taken for each $N$, but 
we have checked that the gross features are not strongly dependent on those
 
\begin{figure}[]
\begin{center}
\includegraphics[width=11cm]{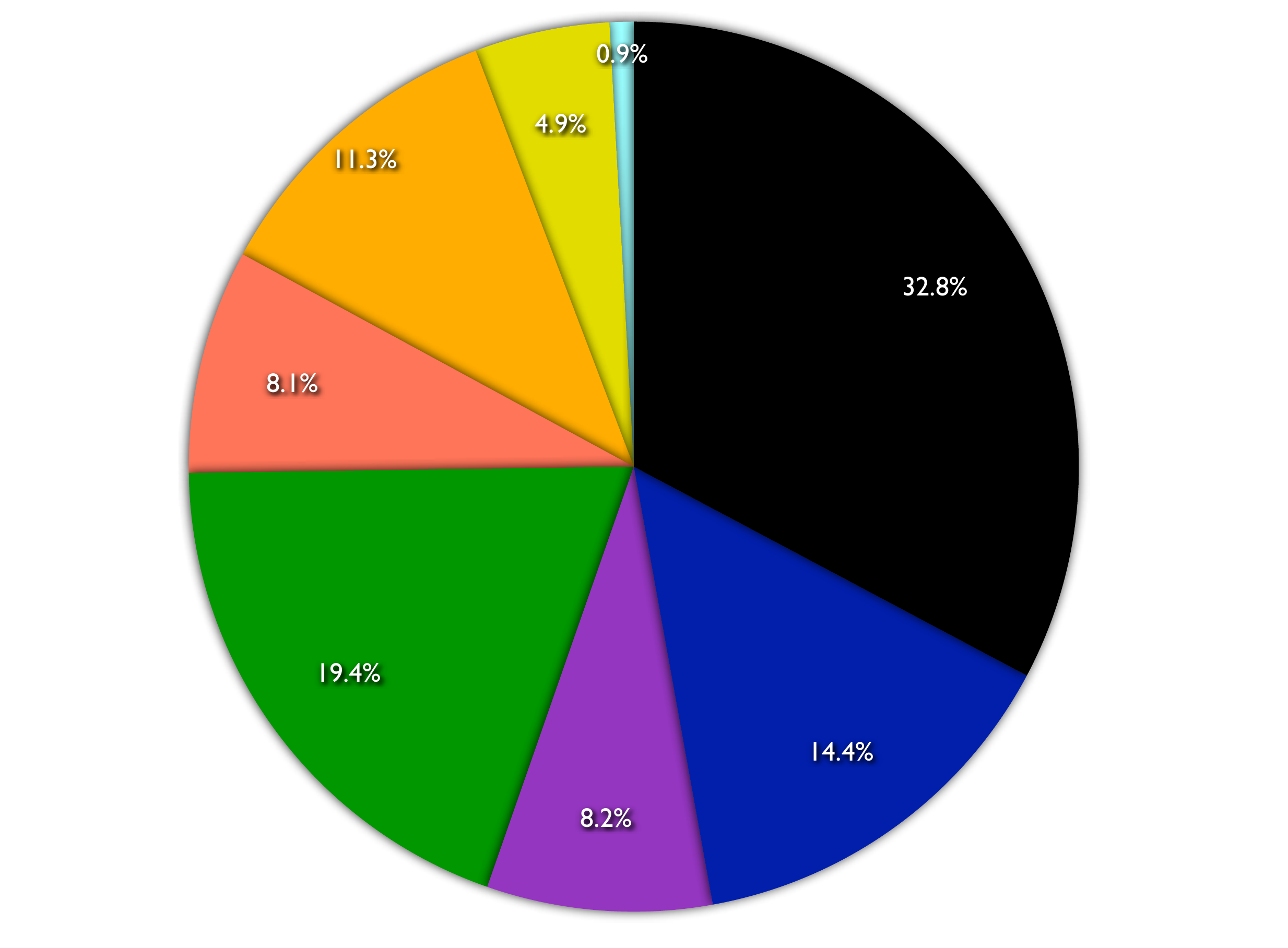}
\caption{Overall distribution of group types.}
\label{AllMipfs}
\end{center}
\end{figure}

In principle there is an exact correspondence between MIPFs of standard and lifted Gepner models: the same MIPF gives rise
to the same group type.   Hence if one randomly select tensor products and MIPFs one would get the same distribution for these
different group types (though the fact that different standard Gepner models may have different numbers of lifts, as well as
differences in randomization introduce some discrepancies). 
This distribution is shown for the lifted Gepner models in Fig. \ref{AllMipfs}.  It is based on $9.4 \times 10^8$ MIPFs, each counted
once, before comparing spectra and identifying identical ones.

\begin{figure}[]
\begin{center}
\includegraphics[width=11cm]{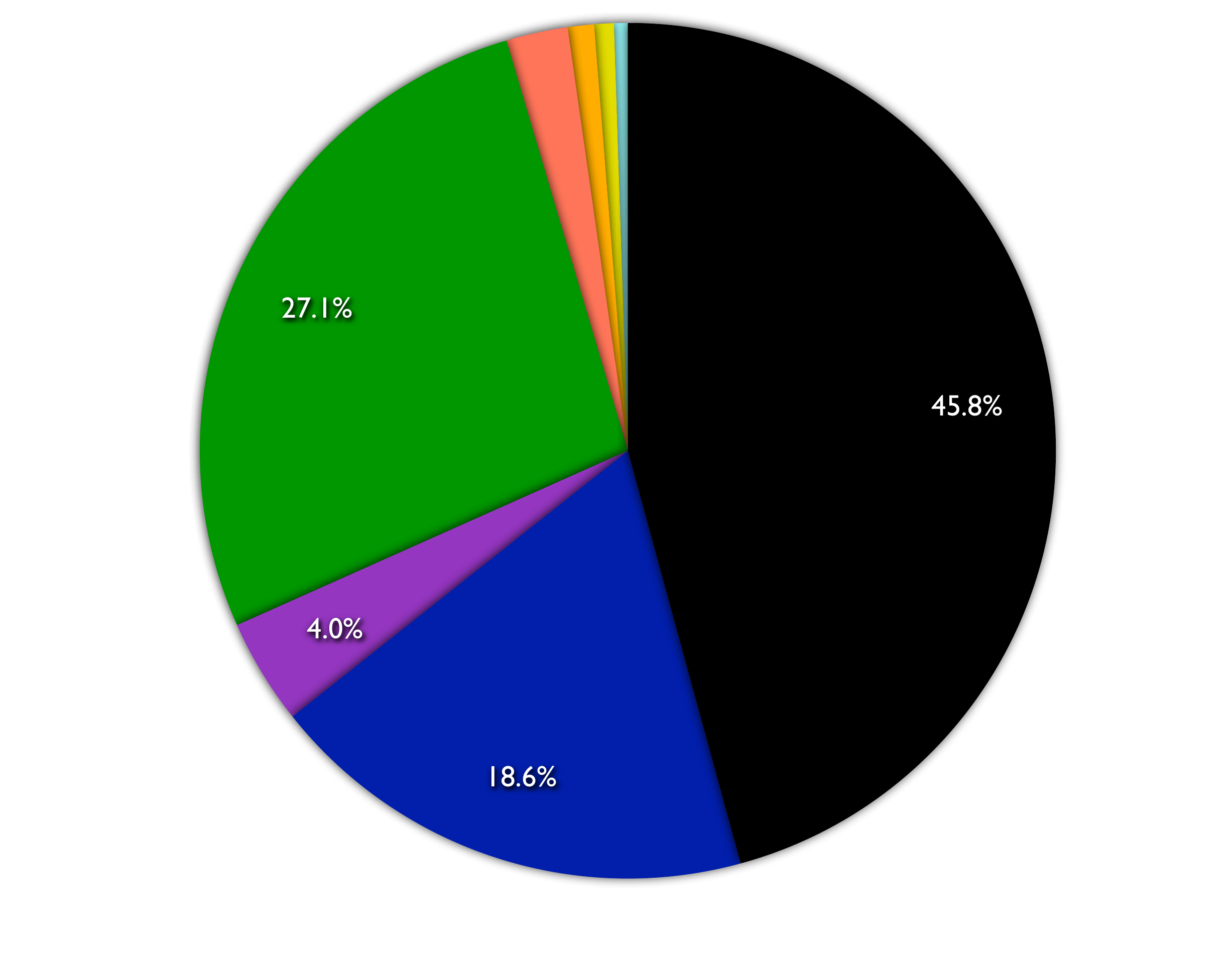}
\caption{Overall distribution of group types for spectra without chiral exotics.}
\label{NonExoticMipfs}
\end{center}
\end{figure}

This distribution contains all spectra, including those with chiral fractionally charged particles. If we require absence
of chiral exotics, the distribution is the one shown in Fig.  \ref{NonExoticMipfs}.  This figure is based on  $6.8 \times 10^8$ MIPFs. 
Note that
requiring absence of chiral exotics reduces the number of spectra by only about $30\%$. 
The reduction is strongest for the spectra with third-integer charges and surprisingly small for those with
sixth-integer  charges. The two half-integer charged types, Pati-Salam and SM, Q=$\frac12$ behave rather differently: the former
is only reduced by less than $10\%$, whereas the latter is reduced by a factor 10. 
In the vast majority of these spectra, fractionally charged exotics are present in the massless spectrum, but they are vector-like. 

\begin{figure}[]
\begin{center}
\includegraphics[width=11cm]{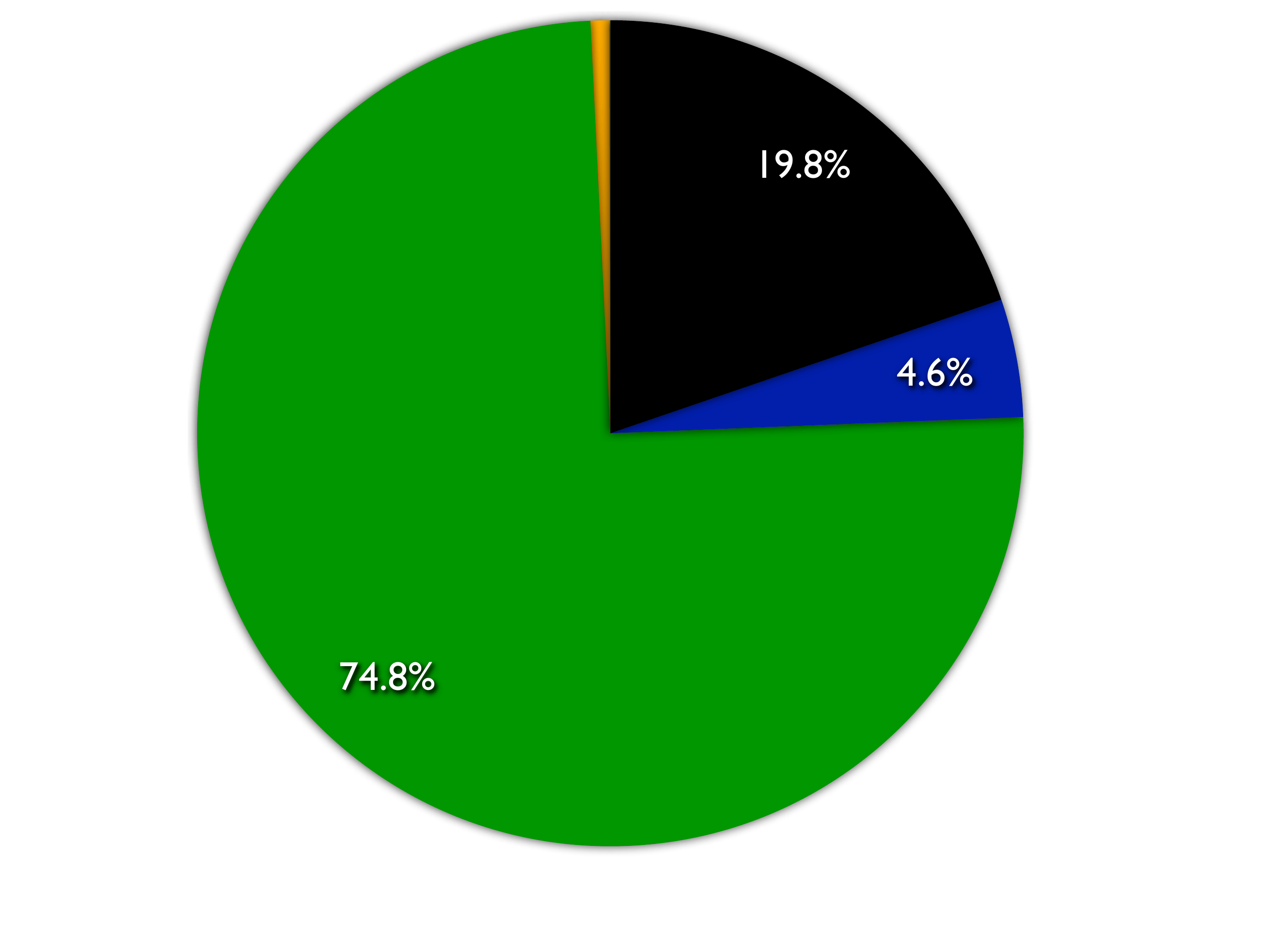}
\caption{Overall distribution of group types  without chiral exotics and at least one family.}
\label{ChiralNonExotics}
\end{center}
\end{figure}

In most cases, the absence of chiral exotics is simply due to the fact that the entire spectrum is non-chiral. So as
a final step we consider only those cases that have at least one chiral family.
Then we get Fig.  \ref{ChiralNonExotics}. Now there
are $1.8 \times 10^8$ spectra left.  The $Q=\frac16$ spectra are reduced to just a few hundred, and are not visible. The $Q=\frac13$
spectra nrs. 1 and 5 occur respectively about 100.000 and 24.000 times, and are also invisible. The only ones that remain with a substantial
frequency are Pati-Salam, SM, Q=$\frac12$, and the two GUT models $SU(5)$ and $SO(10)$. Interestingly, the $SO(10)$ spectra are
reduced much more than the $SU(5)$ models, {\it i.e.} the latter are more often chiral.

\subsection{Fractional charges}

Often the appearance of fractionally charged particles in broken heterotic GUTs is hand-waved away.
Indeed, they may be massive or
confined by additional forces, but the elegance of the original GUTs ($SU(5)$ or $SO(10)$) is lost in either case. More seriously, we
loose a deeper understanding of why the standard model matter comes in the form of the chiral families we observe. 
Without the availability of fractionally charged representations, anomaly cancellation necessarily imposes the 
observed family structure on the chiral spectrum. If in addition one makes the plausible assumption 
that only the chiral spectrum survives at low energies, the observed family structure is nicely
explained. This is indeed true in $SU(5)$ field theory models, but not in heterotic GUTs of the type considered here (it would be true
in heterotic GUTs based on higher level affine Lie algebras, but in that case the set of allowed massless representations 
is not naturally limited to the observed ones).
As soon as fractionally charged matter
is available in principle, there are several other solutions to anomaly cancellation.  
Therefore the observed chiral structure of a standard model family is not understood in this class of heterotic strings, even though it seemed
that with $(16)$'s of $SO(10)$ we were on the right track. If the initial emergence of $(16)$'s of $SO(10)$ is considered a success, by
the same standards the appearance of fractionally charged particles must be seen as a failure. 

It is well-known \cite{Witten:1985xc} that in models of this type (heterotic strings with $SO(10)$ broken to
$SU(3)\times SU(2) \times U(1)$) quarks and leptons are not really unified into a single $(16)$ of $SO(10)$.
They typically originate from several different $(16)$'s in the original manifold. However,
if indeed there were no fractional charges available, one could legitimately argue that in this class of heterotic strings the group-theoretical
fact that a family of quarks and leptons fits in a $(16)$ is understood, even if they do not really belong to the same multiplet. As soon
as fractionally charged anomaly-free multiplets are available {\it in principle}, the best one can still hope for is that models with the observed
family structure dominate the landscape statistically.
Observe that in field-theoretical $SO(10)$ models, upon breaking $SO(10)$, a representation $(16)$ automatically 
decomposes into standard quarks and leptons, whereas in string-theoretic $SO(10)$ models in the heterotic class considered here, this is
not automatic. 
Of course one can impose the bias that only  standard families are allowed in the chiral spectrum, but then on cannot argue anymore
that string theory fully {\it explains} the family structure.

\subsubsection{Chiral fractional charges}

In RCFT constructions like the one we are discussing here, $SU(3)\times SU(2) \times U(1)_{30}$ spectra are sampled without
any further bias towards the standard model family structure (of course the gauge group choice itself {\it is} a bias). 
So in this context
it becomes natural to ask if, despite the existence of other options, the standard model family structure still dominates in a statistical sense.

The ``other options" are models with chiral multiplets that are not of the form of a standard model family. 
We merely count these spectra before rejecting them, in order 
to see to what extent one can still argue that the structure of a standard model family is understood in this class of heterotic strings.

As remarked above, about $71\%$ of randomly generated spectra turned out to be free of chiral exotics, 
whereas for the standard Gepner models this ratio is about $63\%$.  One might be tempted to conclude that 
fractional charges are generically vector-like in these models. 
But this is
a slightly misleading statement for two reasons: on the one hand the set is dominated by $SO(10)$ and $SU(5)$ models, for which this
is automatically true, and on the other hand it is dominated by non-chiral spectra, for which it is also automatically true. 
Another way to ask the question is: which percentage of the chiral spectra with a broken GUT group ({\it i.e.} all types except
$SU(5)$ and $SO(10)$) and a non-anomalous $Y$-charge have a standard family structure with at least one family? 
For standard Gepner
models, the answer is about $20\%$. For lifted Gepner models it is about $4.6\%$. These numbers are not extremely small, but
it is hard to argue that the observed family structure is understood in this class of models. 
Anthropic arguments can be expected to enlarge these ratios (note that we are discussing chiral, and hence necessarily
light or even exactly massless  particles here).
Indeed, we know already that for suitable
parameter values the standard family structure can lead
to the complex nuclear physics and chemistry needed for life, but this is not likely to be the case for the entire set of exotic
 chiral representations
of $SU(3) \times SU(2)\times U(1)$ that emerges from these heterotic strings. Indeed, some of them are lacking some or all
standard quarks and leptons. 
However, analyzing this is unfortunately far beyond anyone's capabilities. 

It would obviously be very interesting to have information about this issue in other constructions (such as orbifolds, Calabi-Yau
compactifications) but usually spectra with chiral exotics are not even considered in these approaches. 
In \cite{Assel:2009xa,Rizos:2010sd,Assel:2010wj} however,
a similar analysis has been done for Pati-Salam models based on free fermions. Their conclusions appear to agree with ours qualitatively:
spectra with chiral exotics dominate those with only standard families by about a factor of five.

\subsubsection{Vector-like fractional charges}

In most of the aforementioned $4.6\%$ of chiral models without chiral exotics, the fractionally charged exotics exist as
vector-like particles. Here we mean vector-like with respect to $SU(3)\times SU(2) \times U(1)$. We did not examine chirality with
respect to other factors in the gauge group. Spectra of this kind are potentially acceptable, provided they acquire a very large mass in a
proper string vacuum, as discussed in section 2.

\subsubsection{Massive fractional charges}

We also found examples where the fractional charges are completely absent from the massless spectrum. In \cite{GatoRivera:2010gv} we
found only very few examples, all of them with zero families; for lifted Gepner models we also have examples with a non-vanishing, though always
even (2, 4, 6, 8, 12 or 24) number of families. These were found only for the Pati-Salam and SM, Q=$\frac12$ group types; in other words, only
if the CFT charge quantum is $\frac12$.  They occur very rarely indeed. To get an idea, let us see the precise numbers for the most 
common one, the Pati-Salam model.  In total, we found about 864.000 distinct\footnote{Note that here each
distinct spectrum is counted only once, whereas the foregoing figures were based on total occurrences} Pati-Salam type models without chiral exotics, of
which about 100.000 were chiral, {\it i.e.} they have a non-vanishing number  of standard model families. About 6000 Pati-Salam models had no
massless exotics at all, but of these 6000 only 143 had a non-zero number of chiral families. For type  SM, Q=$\frac12$ these numbers are respectively
75.000, 36.000, 330 and 36. In both cases, the chance that a chiral model has no fractionally charged exotics at all is about 1 in 1000.
It seems unreasonable to claim that this is the reason we
observe no fractional charges experimentally. 

This same issue has been investigated recently for free fermionic constructions in \cite{Assel:2009xa,Rizos:2010sd,Assel:2010wj}. 
These authors only constructed Pati-Salam models
and found a somewhat smaller rate for the absence of fractional charges, about $10^{-5}$, but did 
find
examples with three families. This demonstrates that there does not exist a no-go theorem against such spectra in the
general class of $(0,2)$ models. Given the fact
that the total number of examples without massless fractional charges we found is of order 100, there is a chance that three family examples 
occur in our set as well at a rate of $10^{-5}$, but
are still hidden in the noise.

The most important conclusion here is that non-exotic models are rare. As explained before, in our opinion 
in RCFT one should aim for absence of chiral exotics, but requiring complete absence of vector-like exotics (fractionally charged or not) is
simply too much to ask for. To put it differently: if one were to reject models because they have some vector-like exotics at the
level of exact RCFT, one risks loosing many examples that would survive perfectly if one perturbs the theory. 

It is known that massless fractional charges can be avoided using Wilson line symmetry breaking in combination with freely acting  
discrete symmetries \cite{Wen:1985qj,Athanasiu:1988uj}; for recent examples see {\it e.g.} \cite{Anderson:2009mh,Blaszczyk:2009in}. 
Let us emphasize that this is not in disagreement with the statement above. The question 
we are trying to answer is how common such non-exotic models are in the full class of examples that have chiral families. 
This question is
not answered by focusing only on the case where massless fractional charges are absent by construction. 
It should be noted that most models
in the ``heterotic mini landscape" studied prior to \cite{Blaszczyk:2009in} (see e.g. \cite{Lebedev:2006kn}) did have exact
spectra with massless fractional charges, whose masses could be lifted by certain terms in the superpotential.  
It would be tremendous progress if a non-phenomenological argument  could be found to explain 
why freely acting discrete symmetries should be favored over other ways to get the standard model from the heterotic string.

Nevertheless, it would be interesting to find out if the examples we found, or those of  \cite{Assel:2009xa}, 
can be understood in terms of some freely acting discrete symmetry (which
is not manifest in our formalism), or if
they are merely statistical fluctuations. The distribution shown in Fig. \ref{Frac}, and which 
will be discussed in section 6, suggests the latter,
since the small peak at zero is not inconsistent with the rest of the distribution (for comparison, the peaks at zero in the first two
plots of Fig. \ref{AbFrac}, to be discussed later, are clearly not statistical fluctuations).

\section{The Number of Families}

\begin{figure}[P]
\begin{center}
\includegraphics[width=13cm]{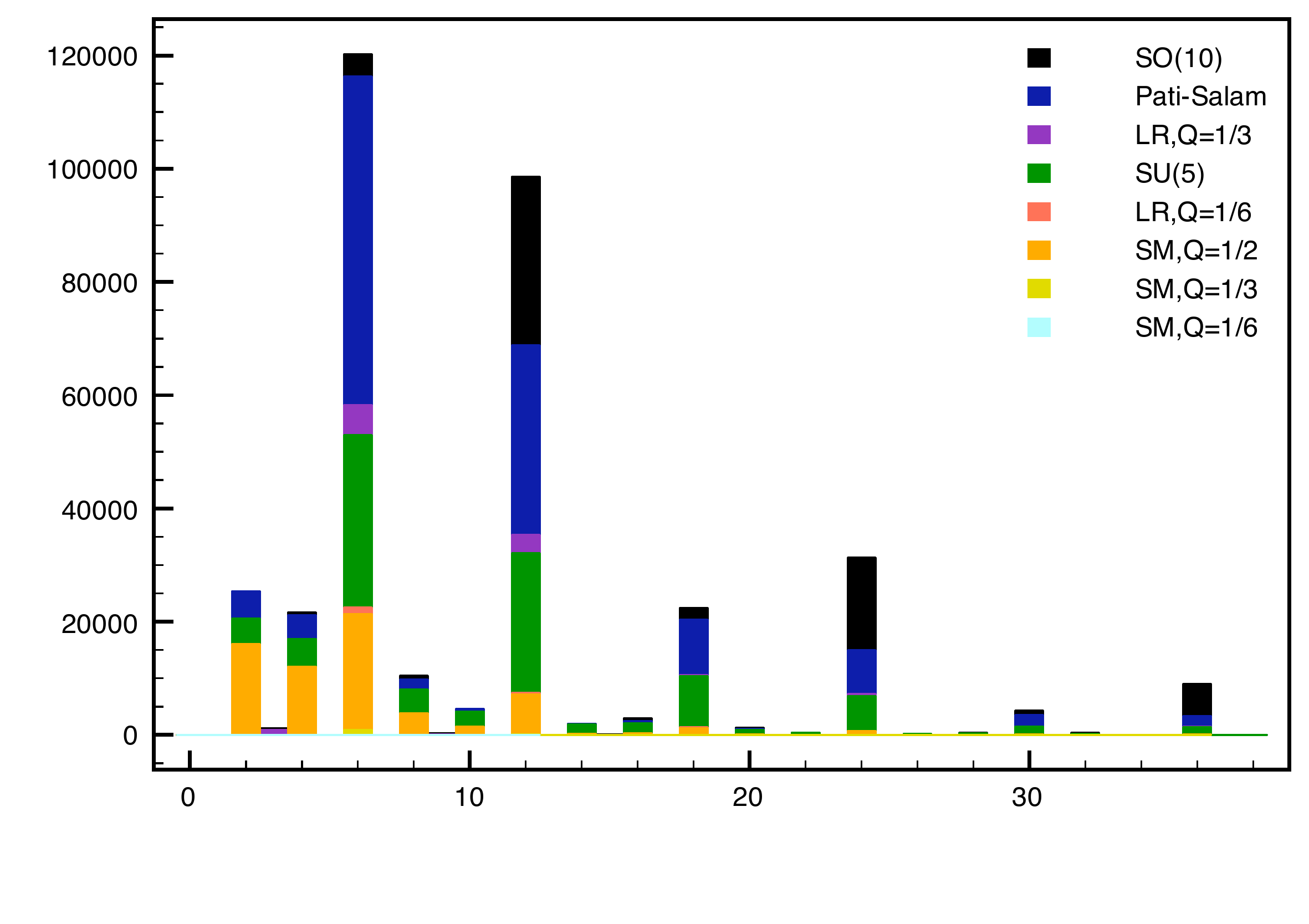}
\caption{Family distribution for standard Gepner models.}
\label{SF}
\end{center}
\end{figure}

\begin{figure}[P]
\begin{center}
\includegraphics[width=13cm]{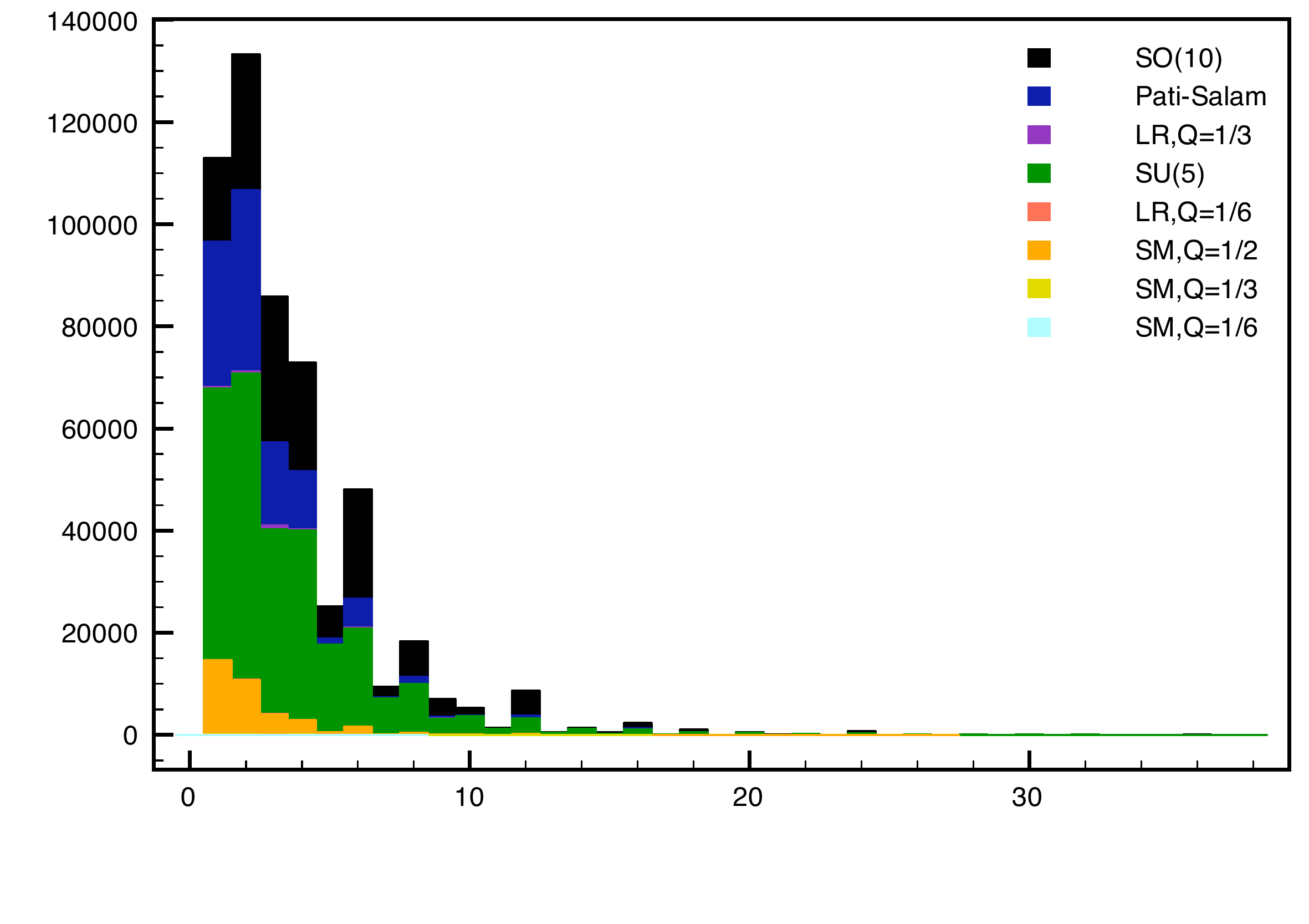}
\caption{Family distribution for lifted  Gepner models.}
\label{SFL}
\end{center}
\end{figure}

For comparison we show in Figs. \ref{SF} and \ref{SFL} the distribution of the number of families for standard Gepner models (168 
minimal model combinations, plus
59 with exceptional invariants of $SU(2)$) and lifted Gepner models (435 combinations). On the vertical axis we show the number of MIPFs.
We distinguish MIPFs on the basis of the numbers of standard model representations Q, U, D, L, E and their conjugates, the total number of
standard model singlets, the total number of fractionally charged particles (not counting their color or weak isospin degrees of freedom,
but counting their dimension in any group that is not part of the standard model), the total number of gauge bosons, the observable part
of the gauge group (gauge bosons that are neutral with respect to $SU(3)\times SU(2)\times U(1)$ are merely counted, but
their mutual interactions are ignored), and any deviations from the expected charge quantization in the massless spectrum 
(for example, if on the basis
of the CFT particles of color singlet charge $\frac16$ are allowed, but in the observed spectrum these particles
all turn out be massive). 

Of course this is not strictly the same as distinguishing MIPFs, since two distinct MIPFs might still produce
the same spectrum, either by coincidence, or by symmetry-related degeneracies (for example the interchange of identical factors in
the tensor product). If there are such degeneracies, it would be preferable to remove them anyway, so no harm is done by identifying
such spectra. Furthermore, such symmetry-related spectra will be identical in all other respects as well. This still leaves the
possibility of coincidental matches of only the massless spectrum. We have reasons to believe these are rare (more about this later), so
that what we are plotting in Figs. \ref{SF} and \ref{SFL} is indeed very close to the number of truly distinct MIPFs. 

Obviously these plots are not true landscape distributions, {\it i.e.} distributions of non-super-symmetric moduli-stabilized string vacua.
Furthermore one may question the relative weights for different MIPFs, and especially the relative weights given to different
tensor products. However, the main point we want to stress in Fig. \ref{SF} is that in standard Gepner models the number three is hard to get,
and in particular very much disfavoured in comparison to two and four, 
confirming the experience of two decades ago. This result remains true regardless of the measure one chooses, unless one can find
an argument explaining why the $(1,16^*,16^*,16^*)$  (giving rise to the only 3-family models in this set)
should be enhanced by a huge factor over anything else.

The main result of the present paper is shown in Fig. \ref{SFL}. We observe that in the lifted Gepner model landscape an approximately 
exponential family distribution is obtained, where the number three is not especially disfavoured with respect to two and is even 
favoured with respect to four.
Indeed, the impression one gets from this plot is  that families with a factor of two and/or three are
{\it enhanced} with respect to the average. Note for example the peaks at 6, 9 and 12 families. 
Also for these models the same caveats regarding true landscape distributions apply. However in this case three family spectra do not just
come from a single tensor product, but from 260 of the 435 combinations, so the issue of their relative weight is less important. 
Of course it is likely that moduli stabilization and supersymmetry breaking affect the shape of this distribution, but we are not aware of
any effect that would single out the number three, either by means of a dip or a peak. It is much more plausible that such effects might
change the slope, and indeed it is  still possible that this suppresses three families with respect to two by a huge amount, but the optimistic
interpretation of our results is that in the heterotic landscape, as soon as one moves away from the $(2,2)$ models, the number three
is not especially difficult to obtain.

Further evidence for the special role of the numbers two and three comes from studying the greatest common divisor 
(denoted $\Delta$) of the number
of families for all the MIPFs of each lifted $N=2$ tensor product. The results are listed in Table \ref{DeltaDist}. 
Most lifted Gepner models have a
value $\Delta=1$, but 0, 2, 3, 4, and 6 also occur.  By contrast, for standard Gepner models, the values of $\Delta$ we found were
$0,2,3,6$ and $12$, with 1 completely absent and 3 occurring just once.  The most noticeable difference is the large 
number of cases we get now with $\Delta=1$. There are no cases with $\Delta=5$ or any number larger than 6. 
We do get spectra with 5, 7, or 11, $\ldots$ families from the
$\Delta=1$ models; the first number that we did not encounter at all is 31.  

The distribution of families was also studied recently in the context of heterotic strings using free fermions \cite{Assel:2010wj}.
These authors also find a roughly exponential fall-off with the number of families, but with a steeper slope of about a factor three
per family. Even though in that case three is more suppressed, one could still conclude that three family spectra are not especially rare. 
An interesting difference with
our results is the complete absence of spectra with seven families (or any larger odd number). This must perhaps be viewed as
an artifact of free fermionic constructions.

\begin{table}[h]
\begin{center}
\vskip .7truecm
\begin{tabular}{|c||c|}
\hline
\hline
$\Delta$ & Number \\
\hline
0 & 94    \\
1 & 199    \\
2 & 56   \\
3 & 61     \\
4 & 8    \\
5 & 0    \\
6 & 17    \\
\hline
\end{tabular}
\vskip .7truecm
\caption{The number of lifted Gepner models with a given family quantum $\Delta$.}
\label{DeltaDist}
\end{center}
\end{table}

Our conclusion is quite different from what was observed for orientifolds in \cite{Dijkstra:2004cc} and \cite{Gmeiner:2005vz}. In this
case the number of families drops off much faster, and there is a clear additional suppression of odd numbers with respect to even ones,
which unfortunately is not understood. Consequently, three is suppressed with respect to two and four by two to three orders
of magnitude. It is impossible to tell whether the heterotic results or the orientifolds results are more typical for the string 
landscape as a whole, but at least there is a better chance now that the difficulties with obtaining three families can be 
attributed to artifacts of the chosen method.

\begin{figure}[P]
\begin{center}
\includegraphics[width=13cm]{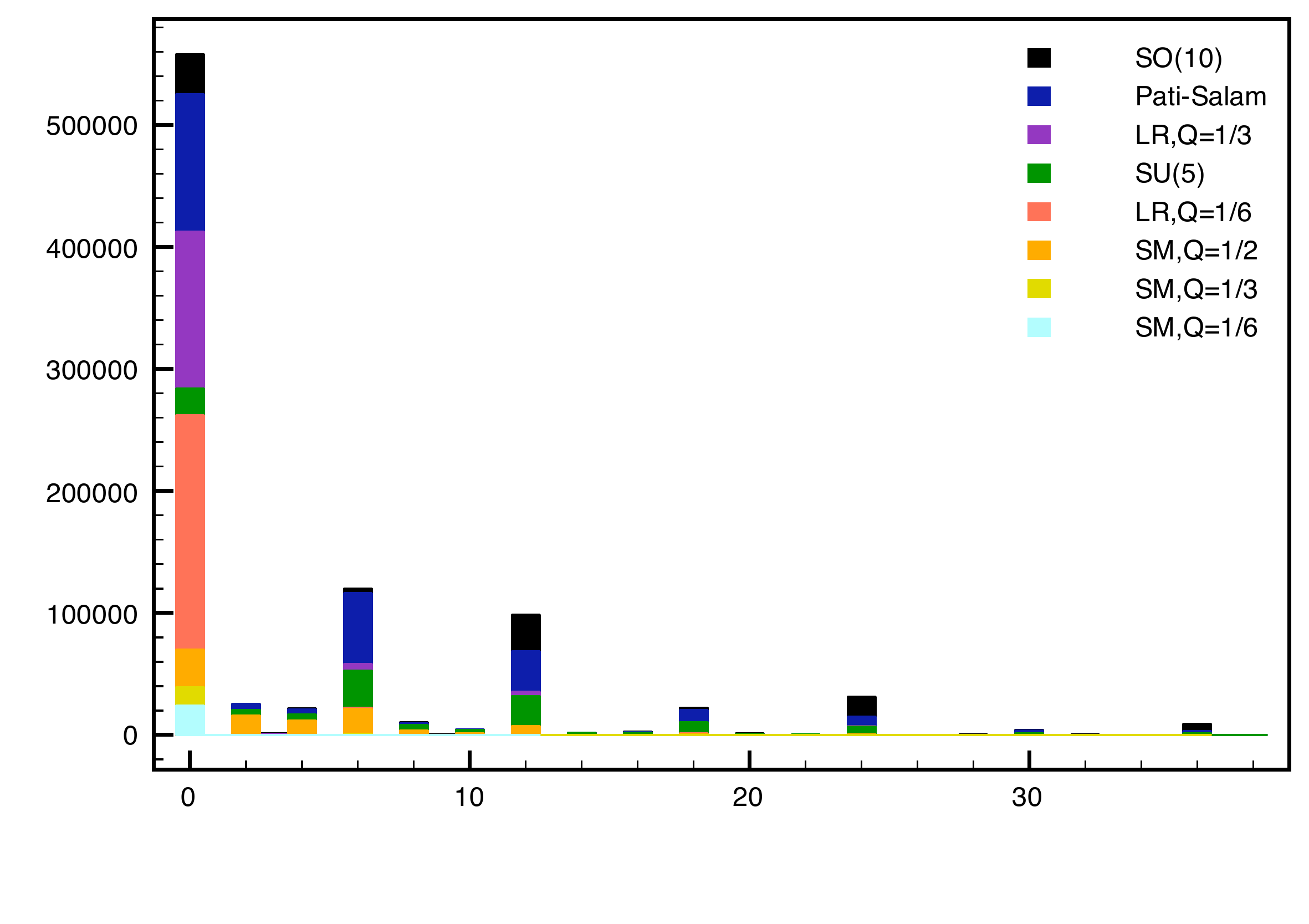}
\caption{Family distribution for standard Gepner models, including zero families}
\label{SFZ}
\end{center}
\end{figure}

\begin{figure}[P]
\begin{center}
\includegraphics[width=13cm]{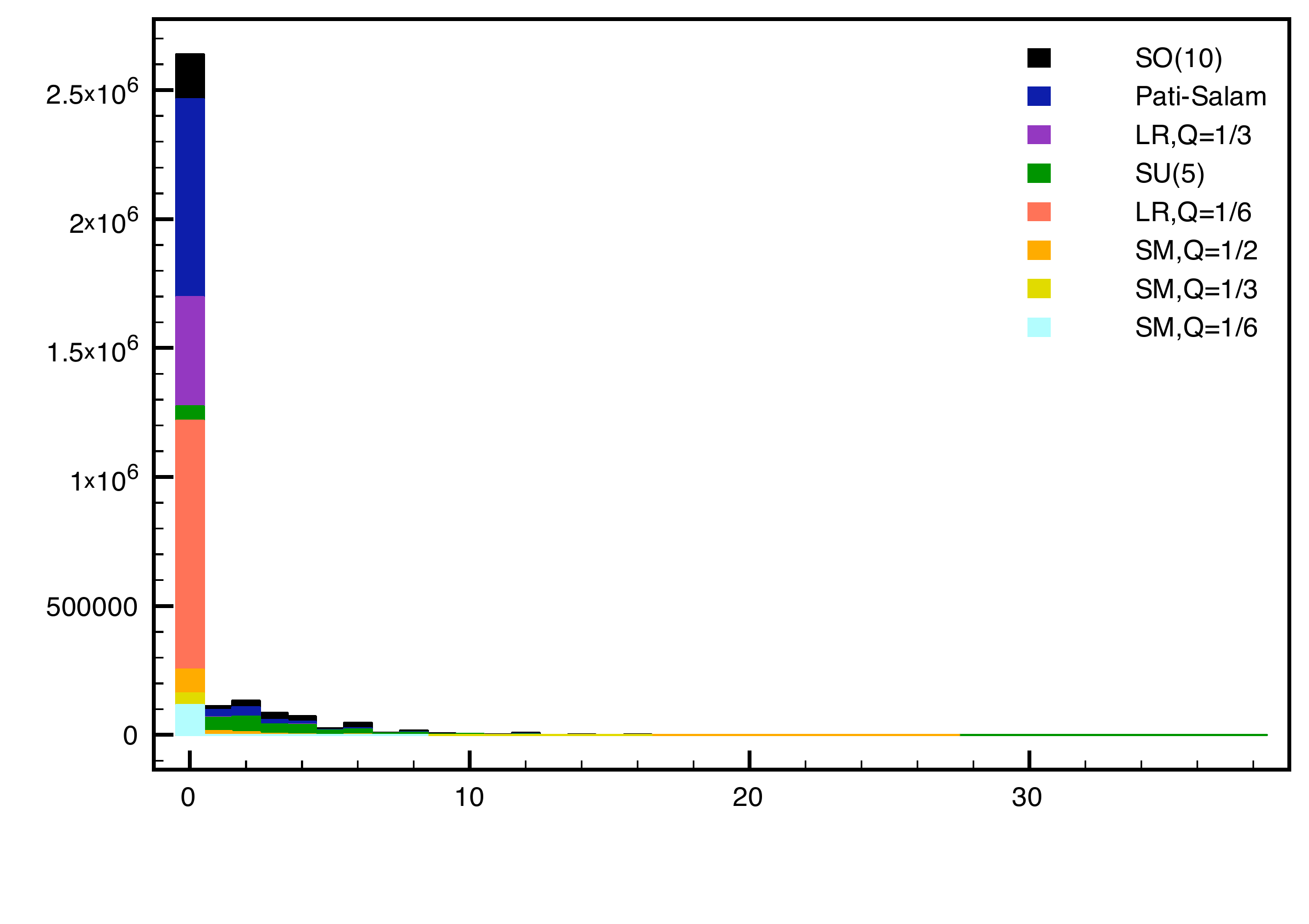}
\caption{Family distribution for lifted Gepner models, including zero families.}
\label{SFLZ}
\end{center}
\end{figure}

Note that Fig. \ref{SFL} is dominated by just four group types, the phenomenologically useless (because unbroken) GUT groups
$SU(5)$ and $SO(10)$, the Pati-Salam model, and the unextended standard model with charge quantization in half-integer units. 
The other four types are also present, as was already discussed in the previous section, but there are too few of them to be visible. 
To make them appear we can add the data for zero family models. This is shown for standard Gepner models in Fig. \ref{SFZ} and 
for lifted Gepner models in Fig. \ref{SFLZ}. Especially in the latter case this yields a huge bar at zero that dominates everything. Universes
with zero families thus seem to dominate the ensemble, but this is not a problem, since they are anthropically unacceptable.

In comparison with the results of \cite{Assel:2010wj} we get a much larger peak at zero families. This can be explained, at least partly,
as follows. Our plots are in terms of the group types introduced in the previous section. These group types define the subgroup
of $SO(10)$ that is realized, but we do not put any constraints on the extension of these groups into the internal
sector of the theory, whereas the authors of \cite{Assel:2010wj} require that there be no extension beyond the group they consider,
the Pati-Salam model. Some of those extensions will render the gauge group non-chiral, and therefore can only
contribute to the zero-family peak in the distribution, producing an abnormal enhancement of that peak.

Note that in Figs. \ref{SF} $\ldots$ \ref{SFLZ} on the vertical axis we have plotted the number of distinct MIPFs, and not the absolute
occurrence frequency.  This explains why the relative distribution of group types appears different than in the pie charts shown in the
previous section. If MIPFs are selected at random, the unbroken $SU(5)$ and $SO(10)$ cases tend to come out far more often, and
hence they dominate the set. The reason for using total occurrence frequency rather than the total number of distinct MIPFs in the
foregoing section was explained in \cite{GatoRivera:2010gv}: we do not distinguish MIPFs for spectra with chiral exotics.

In Table \ref{HWLSummary} we show all lifted Gepner models with a non-vanishing value for $\Delta$. In the first column
the combination of levels is shown, with a hat indicating which factor is lifted. For some values of the level $k$ two lifts
are possible. The second one is indicated by a tilde instead of a hat. Column 2 shows the value of $\Delta$ and 
column 3 the maximal number of families for that model.
Column 4 lists the number of distinct 3-family spectra we obtained, and column 5 lists the total number of distinct $N$ family
spectra, with $N \geq 1$. In both of these columns we have modded out mirror symmetry, {\it i.e.} a spectrum and its mirror pair
is just counted once.

As explained in \cite{GatoRivera:2010gv} the full set of spectra we obtain is mirror symmetric. This is a statement about RCFT, 
and does not necessarily
imply anything about an underlying geometry. Typically, in order to encounter a mirror partner of a given spectrum it is necessary
to add one simple current twist. Therefore, if we go to arbitrary numbers of generators of the simple current subgroup, we will
eventually find all mirror pairs. For practical reasons, we have limited this number to 4. By checking if the set of spectra is closed under
mirror symmetry we can get a rough idea about the completeness of the set. For this reason we have indicated in the last column
which percentage of the total sets of MIPFs is lacking a mirror partner. As expected, this number tends to be largest for tensor combinations
with many factors, because here the limitation to four generators is most restrictive.  

\LTcapwidth=14truecm
\begin{center}
\vskip .7truecm
\begin{longtable}{|c||c|c|c|c|c|}\caption{{\bf{Results for lifted Gepner models}}}\\
\hline
 \multicolumn{1}{|c||}{model}
& \multicolumn{1}{c|}{$\Delta$}
& \multicolumn{1}{l|}{Max. }
& \multicolumn{1}{c|}{3 family}
& \multicolumn{1}{c|}{$N$ fam. ($N>0$)}  
& \multicolumn{1}{c|}{Missing Mirrors } \\ 
\hline
\endfirsthead
\multicolumn{6}{c}%
{{\bfseries \tablename\ \thetable{} {\rm-- continued from previous page}}} \\
\hline 
 \multicolumn{1}{|c||}{model}
& \multicolumn{1}{c|}{$\Delta$}
& \multicolumn{1}{l|}{Max. }
& \multicolumn{1}{c|}{3 family}
& \multicolumn{1}{c|}{$N$ fam. ($N>0$)}  
& \multicolumn{1}{c|}{Missing Mirrors } \\ 
\hline
\endhead
\hline \multicolumn{6}{|r|}{{Continued on next page}} \\ \hline
\endfoot
\hline \hline
\endlastfoot\hline
\label{HWLSummary}
$(\widehat{1},10,11,154)$ & 6 & 6 & 0 & 4 &    0.00\% \\ 
$(1,10,\widehat{11},154)$ & 6 & 6 & 0 & 8 &    0.00\% \\ 
$(1,10,\widehat{12},82)$ & 1 & 8 & 10 & 69 &    0.00\% \\ 
$(\widehat{1},10,13,58)$ & 1 & 18 & 189 & 563 &    0.71\% \\ 
$(1,10,\widehat{13},58)$ & 1 & 64 & 11 & 295 &    0.34\% \\ 
$(1,10,13,\widehat{58})$ & 1 & 42 & 7 & 103 &    0.00\% \\ 
$(\widehat{1},10,14,46)$ & 3 & 12 & 126 & 172 &    8.14\% \\ 
$(1,10,\widehat{14},46)$ & 1 & 14 & 96 & 648 &   12.50\% \\ 
$(\widehat{1},10,18,28)$ & 1 & 17 & 126 & 560 &    6.79\% \\ 
$(\widehat{1},10,19,26)$ & 3 & 6 & 8 & 12 &    0.00\% \\ 
$(1,10,\widehat{19},26)$ & 3 & 15 & 2 & 10 &    0.00\% \\ 
$(1,10,19,\widehat{26})$ & 1 & 12 & 1 & 9 &    0.00\% \\ 
$(\widehat{1},10,22,22)$ & 3 & 24 & 117 & 215 &   24.19\% \\ 
$(\widehat{1},1,1,1,10,10)$ & 3 & 30 & 757 & 1194 &   31.49\% \\ 
$(\widehat{1},1,1,1,1,1,1,1,1)$ & 3 & 33 & 304 & 584 &   40.07\% \\ 
$(1,1,1,1,1,1,1,\widehat{4})$ & 2 & 24 & 0 & 437 &   33.64\% \\ 
$(\widehat{1},1,1,1,1,1,1,4)$ & 3 & 33 & 359 & 670 &   27.76\% \\ 
$(1,1,1,1,1,\widehat{2},10)$ & 1 & 24 & 26 & 226 &   15.04\% \\ 
$(\widehat{1},1,1,1,1,2,10)$ & 3 & 30 & 98 & 343 &   26.24\% \\ 
$(1,1,1,1,1,\widehat{4},4)$ & 1 & 24 & 344 & 2986 &   37.51\% \\ 
$(\widehat{1},1,1,1,1,4,4)$ & 3 & 33 & 1827 & 3054 &   27.96\% \\ 
$(\widehat{1},11,11,76)$ & 6 & 6 & 0 & 2 &    0.00\% \\ 
$(1,\widehat{11},11,76)$ & 3 & 6 & 4 & 7 &    0.00\% \\ 
$(1,11,\widehat{11},76)$ & 2 & 8 & 0 & 6 &    0.00\% \\ 
$(1,1,1,1,2,2,\widehat{4})$ & 1 & 16 & 0 & 100 &   31.00\% \\ 
$(\widehat{1},1,1,1,2,2,4)$ & 3 & 24 & 81 & 186 &   20.43\% \\ 
$(1,1,1,1,\widehat{2},2,4)$ & 1 & 24 & 13 & 175 &   29.14\% \\ 
$(\widehat{1},1,1,1,5,40)$ & 3 & 18 & 12 & 25 &   36.00\% \\ 
$(1,1,1,1,\widehat{5},40)$ & 2 & 8 & 0 & 21 &   23.81\% \\ 
$(\widehat{1},1,1,1,6,22)$ & 3 & 24 & 64 & 143 &   14.69\% \\ 
$(1,1,1,1,\widehat{6},22)$ & 1 & 16 & 0 & 81 &   13.58\% \\ 
$(\widehat{1},1,1,1,7,16)$ & 3 & 27 & 303 & 523 &   29.45\% \\ 
$(\widehat{1},1,1,1,8,13)$ & 3 & 18 & 7 & 37 &    8.11\% \\ 
$(1,1,1,1,\widehat{8},13)$ & 2 & 16 & 0 & 19 &    0.00\% \\ 
$(\widehat{1},1,1,2,4,10)$ & 3 & 30 & 706 & 1377 &   27.09\% \\ 
$(1,1,1,\widehat{2},4,10)$ & 1 & 24 & 315 & 1542 &   25.42\% \\ 
$(1,1,1,2,\widehat{4},10)$ & 1 & 24 & 114 & 1245 &   22.25\% \\ 
$(\widehat{1},1,1,4,4,4)$ & 3 & 33 & 4117 & 6732 &   28.58\% \\ 
$(1,1,1,\widehat{4},4,4)$ & 1 & 32 & 1044 & 9903 &   22.56\% \\ 
$(1,\widehat{12},12,40)$ & 1 & 18 & 21 & 123 &    0.00\% \\ 
$(\widehat{1},1,2,12,82)$ & 3 & 18 & 28 & 55 &    0.00\% \\ 
$(1,1,\widehat{2},12,82)$ & 1 & 14 & 16 & 70 &    0.00\% \\ 
$(1,1,2,\widehat{12},82)$ & 2 & 8 & 0 & 15 &    0.00\% \\ 
$(\widehat{1},1,2,13,58)$ & 1 & 18 & 42 & 156 &    4.49\% \\ 
$(1,1,\widehat{2},13,58)$ & 1 & 14 & 22 & 160 &    3.12\% \\ 
$(1,1,2,\widehat{13},58)$ & 1 & 18 & 1 & 60 &    1.67\% \\ 
$(1,1,2,13,\widehat{58})$ & 2 & 32 & 0 & 19 &    5.26\% \\ 
$(\widehat{1},1,2,14,46)$ & 3 & 15 & 144 & 211 &    6.16\% \\ 
$(1,1,\widehat{2},14,46)$ & 1 & 12 & 42 & 174 &    0.00\% \\ 
$(1,1,2,\widehat{14},46)$ & 1 & 12 & 9 & 195 &    2.05\% \\ 
$(\widehat{1},1,2,16,34)$ & 3 & 21 & 368 & 734 &   19.75\% \\ 
$(1,1,\widehat{2},16,34)$ & 1 & 18 & 73 & 816 &   12.50\% \\ 
$(\widehat{1},1,2,18,28)$ & 1 & 12 & 98 & 247 &    1.62\% \\ 
$(1,1,\widehat{2},18,28)$ & 1 & 12 & 42 & 204 &    1.96\% \\ 
$(\widehat{1},12,19,19)$ & 3 & 6 & 14 & 18 &    0.00\% \\ 
$(1,\widehat{12},19,19)$ & 1 & 14 & 0 & 9 &    0.00\% \\ 
$(1,12,\widehat{19},19)$ & 6 & 6 & 0 & 1 &    0.00\% \\ 
$(\widehat{1},1,2,2,2,10)$ & 3 & 24 & 219 & 458 &    5.68\% \\ 
$(1,1,\widehat{2},2,2,10)$ & 1 & 36 & 263 & 1963 &   21.40\% \\ 
$(\widehat{1},1,2,22,22)$ & 3 & 30 & 200 & 481 &   13.31\% \\ 
$(1,1,\widehat{2},22,22)$ & 1 & 24 & 43 & 471 &   11.46\% \\ 
$(\widehat{1},1,2,2,4,4)$ & 3 & 24 & 285 & 771 &   13.62\% \\ 
$(1,1,\widehat{2},2,4,4)$ & 1 & 24 & 24 & 854 &   17.80\% \\ 
$(1,1,2,2,\widehat{4},4)$ & 1 & 32 & 81 & 1555 &   20.19\% \\ 
$(\widehat{1},1,3,13,13)$ & 1 & 24 & 220 & 569 &   20.74\% \\ 
$(1,1,\widehat{3},13,13)$ & 1 & 48 & 28 & 580 &    5.52\% \\ 
$(1,1,3,\widehat{13},13)$ & 1 & 36 & 47 & 475 &    6.53\% \\ 
$(\widehat{1},13,13,28)$ & 1 & 34 & 335 & 1093 &    5.67\% \\ 
$(1,\widehat{13},13,28)$ & 1 & 80 & 145 & 1184 &    2.20\% \\ 
$(\widehat{1},13,18,18)$ & 1 & 24 & 37 & 240 &    0.00\% \\ 
$(1,\widehat{13},18,18)$ & 1 & 48 & 10 & 137 &    0.00\% \\ 
$(\widehat{1},1,3,7,43)$ & 1 & 18 & 51 & 94 &    9.57\% \\ 
$(1,1,\widehat{3},7,43)$ & 1 & 24 & 6 & 114 &    8.77\% \\ 
$(\widehat{1},1,3,8,28)$ & 1 & 22 & 251 & 922 &    7.70\% \\ 
$(1,1,\widehat{3},8,28)$ & 1 & 48 & 145 & 1279 &   15.72\% \\ 
$(1,1,3,\widehat{8},28)$ & 1 & 32 & 42 & 334 &    4.49\% \\ 
$(\widehat{1},1,4,10,10)$ & 3 & 30 & 2203 & 3775 &   11.42\% \\ 
$(1,1,\widehat{4},10,10)$ & 1 & 24 & 269 & 2842 &   36.17\% \\ 
$(\widehat{1},14,14,22)$ & 3 & 24 & 34 & 79 &    3.80\% \\ 
$(1,\widehat{14},14,22)$ & 1 & 12 & 53 & 379 &   15.57\% \\ 
$(1,14,\widehat{14},22)$ & 1 & 13 & 60 & 305 &   20.00\% \\ 
$(\widehat{1},1,4,5,40)$ & 3 & 18 & 18 & 48 &    0.00\% \\ 
$(1,1,\widehat{4},5,40)$ & 2 & 16 & 0 & 35 &   20.00\% \\ 
$(1,1,4,\widehat{5},40)$ & 2 & 12 & 0 & 22 &    4.55\% \\ 
$(\widehat{1},1,4,6,22)$ & 3 & 24 & 504 & 1016 &    9.55\% \\ 
$(1,1,\widehat{4},6,22)$ & 1 & 24 & 78 & 1009 &   12.88\% \\ 
$(1,1,4,\widehat{6},22)$ & 1 & 48 & 31 & 837 &    9.44\% \\ 
$(\widehat{1},1,4,7,16)$ & 3 & 27 & 900 & 1476 &   21.00\% \\ 
$(1,1,\widehat{4},7,16)$ & 1 & 20 & 196 & 1596 &   10.78\% \\ 
$(\widehat{1},1,4,8,13)$ & 1 & 18 & 43 & 125 &    2.40\% \\ 
$(1,1,\widehat{4},8,13)$ & 1 & 16 & 8 & 107 &    0.93\% \\ 
$(1,1,4,\widehat{8},13)$ & 2 & 16 & 0 & 31 &    0.00\% \\ 
$(1,1,4,8,\widehat{13})$ & 1 & 18 & 3 & 20 &    0.00\% \\ 
$(\widehat{1},1,5,5,19)$ & 3 & 12 & 5 & 12 &    0.00\% \\ 
$(1,1,\widehat{5},5,19)$ & 1 & 8 & 7 & 9 &    0.00\% \\ 
$(1,1,5,\widehat{5},19)$ & 3 & 6 & 7 & 8 &    0.00\% \\ 
$(\widehat{1},1,6,6,10)$ & 3 & 6 & 48 & 71 &    0.00\% \\ 
$(1,1,\widehat{6},6,10)$ & 1 & 8 & 31 & 121 &    0.00\% \\ 
$(\widehat{1},1,7,7,7)$ & 3 & 27 & 246 & 315 &    1.90\% \\ 
$(\widehat{1},2,2,10,10)$ & 3 & 24 & 620 & 1060 &    5.38\% \\ 
$(1,\widehat{2},2,10,10)$ & 1 & 25 & 764 & 5365 &   25.55\% \\ 
$(1,2,2,2,2,\widehat{4})$ & 2 & 32 & 0 & 293 &   15.02\% \\ 
$(\widehat{1},2,2,2,2,4)$ & 6 & 24 & 0 & 116 &   18.97\% \\ 
$(1,\widehat{2},2,2,2,4)$ & 2 & 32 & 0 & 1499 &   11.41\% \\ 
$(\widehat{1},2,2,6,22)$ & 3 & 24 & 136 & 423 &    9.46\% \\ 
$(1,\widehat{2},2,6,22)$ & 1 & 18 & 204 & 1717 &   21.26\% \\ 
$(1,2,2,\widehat{6},22)$ & 1 & 48 & 75 & 1010 &   16.93\% \\ 
$(\widehat{1},2,2,7,16)$ & 3 & 12 & 3 & 17 &    0.00\% \\ 
$(1,\widehat{2},2,7,16)$ & 1 & 12 & 0 & 61 &    6.56\% \\ 
$(\widehat{1},2,3,3,58)$ & 1 & 12 & 57 & 203 &    2.96\% \\ 
$(1,\widehat{2},3,3,58)$ & 1 & 16 & 31 & 245 &    2.04\% \\ 
$(1,2,\widehat{3},3,58)$ & 1 & 12 & 54 & 493 &   10.34\% \\ 
$(1,2,3,3,\widehat{58})$ & 1 & 24 & 0 & 52 &    0.00\% \\ 
$(\widehat{1},2,3,4,18)$ & 2 & 12 & 0 & 24 &    0.00\% \\ 
$(1,\widehat{2},3,4,18)$ & 2 & 24 & 0 & 76 &    0.00\% \\ 
$(1,2,\widehat{3},4,18)$ & 2 & 24 & 0 & 103 &    3.88\% \\ 
$(1,2,3,\widehat{4},18)$ & 2 & 16 & 0 & 51 &    0.00\% \\ 
$(1,\widehat{2},4,4,10)$ & 1 & 24 & 419 & 3016 &   15.02\% \\ 
$(1,2,\widehat{4},4,10)$ & 1 & 20 & 451 & 4097 &   22.19\% \\ 
$(\widehat{1},2,4,6,6)$ & 6 & 12 & 0 & 18 &   11.11\% \\ 
$(1,\widehat{2},4,6,6)$ & 2 & 24 & 0 & 172 &    8.14\% \\ 
$(1,2,\widehat{4},6,6)$ & 4 & 16 & 0 & 35 &    8.57\% \\ 
$(1,2,4,\widehat{6},6)$ & 2 & 48 & 0 & 305 &   14.75\% \\ 
$(\widehat{1},3,3,3,13)$ & 1 & 18 & 125 & 692 &    3.18\% \\ 
$(1,\widehat{3},3,3,13)$ & 1 & 36 & 444 & 3383 &   12.33\% \\ 
$(1,3,3,3,\widehat{13})$ & 1 & 36 & 41 & 410 &    0.73\% \\ 
$(\widehat{1},3,3,4,8)$ & 2 & 14 & 0 & 63 &    0.00\% \\ 
$(1,\widehat{3},3,4,8)$ & 1 & 24 & 18 & 411 &   12.65\% \\ 
$(1,3,3,\widehat{4},8)$ & 2 & 18 & 0 & 122 &    0.82\% \\ 
$(1,3,3,4,\widehat{8})$ & 1 & 18 & 3 & 51 &    1.96\% \\ 
$(1,\widehat{4},4,4,4)$ & 1 & 24 & 947 & 8018 &   12.91\% \\ 
$(\widehat{1},5,42,922)$ & 6 & 24 & 0 & 8 &    0.00\% \\ 
$(1,\widehat{5},42,922)$ & 1 & 28 & 0 & 15 &    0.00\% \\ 
$(1,5,\widehat{42},922)$ & 2 & 44 & 0 & 18 &    0.00\% \\ 
$(\widehat{1},5,43,628)$ & 1 & 18 & 5 & 19 &    5.26\% \\ 
$(1,\widehat{5},43,628)$ & 1 & 24 & 6 & 27 &    7.41\% \\ 
$(\widehat{1},5,46,334)$ & 3 & 24 & 5 & 50 &    2.00\% \\ 
$(1,\widehat{5},46,334)$ & 1 & 28 & 8 & 67 &    0.00\% \\ 
$(\widehat{1},5,47,292)$ & 6 & 24 & 0 & 8 &    0.00\% \\ 
$(1,\widehat{5},47,292)$ & 1 & 28 & 0 & 7 &    0.00\% \\ 
$(1,5,\widehat{47},292)$ & 3 & 12 & 3 & 5 &    0.00\% \\ 
$(\widehat{1},5,52,187)$ & 6 & 12 & 0 & 5 &    0.00\% \\ 
$(1,\widehat{5},52,187)$ & 1 & 16 & 0 & 15 &   20.00\% \\ 
$(\widehat{1},5,54,166)$ & 3 & 30 & 0 & 51 &    0.00\% \\ 
$(1,\widehat{5},54,166)$ & 1 & 36 & 13 & 68 &    1.47\% \\ 
$(1,5,\widehat{54},166)$ & 1 & 20 & 0 & 43 &    0.00\% \\ 
$(\widehat{1},5,58,138)$ & 1 & 12 & 2 & 26 &    0.00\% \\ 
$(1,\widehat{5},58,138)$ & 1 & 16 & 6 & 33 &    0.00\% \\ 
$(1,5,\widehat{58},138)$ & 2 & 48 & 0 & 9 &    0.00\% \\ 
$(\widehat{1},5,61,124)$ & 3 & 36 & 5 & 23 &    8.70\% \\ 
$(1,\widehat{5},61,124)$ & 1 & 42 & 1 & 46 &   23.91\% \\ 
$(\widehat{1},5,68,103)$ & 1 & 18 & 18 & 40 &   12.50\% \\ 
$(1,\widehat{5},68,103)$ & 1 & 22 & 2 & 34 &    0.00\% \\ 
$(\widehat{1},5,82,82)$ & 3 & 48 & 6 & 53 &    0.00\% \\ 
$(1,\widehat{5},82,82)$ & 1 & 56 & 22 & 81 &    1.23\% \\ 
$(\widehat{1},6,23,598)$ & 1 & 2 & 0 & 12 &    8.33\% \\ 
$(1,\widehat{6},23,598)$ & 2 & 8 & 0 & 16 &    0.00\% \\ 
$(1,6,\widehat{23},598)$ & 2 & 12 & 0 & 50 &    8.00\% \\ 
$(\widehat{1},6,24,310)$ & 6 & 12 & 0 & 11 &   18.18\% \\ 
$(1,\widehat{6},24,310)$ & 1 & 32 & 10 & 104 &    0.96\% \\ 
$(\widehat{1},6,25,214)$ & 3 & 12 & 12 & 24 &    0.00\% \\ 
$(1,\widehat{6},25,214)$ & 2 & 32 & 0 & 57 &    8.77\% \\ 
$(\widehat{1},6,26,166)$ & 6 & 6 & 0 & 14 &    0.00\% \\ 
$(1,\widehat{6},26,166)$ & 2 & 24 & 0 & 128 &    0.78\% \\ 
$(1,6,\widehat{26},166)$ & 1 & 10 & 6 & 70 &    2.86\% \\ 
$(\widehat{1},6,28,118)$ & 1 & 20 & 236 & 1003 &   21.14\% \\ 
$(1,\widehat{6},28,118)$ & 1 & 40 & 174 & 1531 &   24.62\% \\ 
$(\widehat{1},6,30,94)$ & 3 & 12 & 88 & 168 &   26.19\% \\ 
$(1,\widehat{6},30,94)$ & 1 & 40 & 43 & 483 &   19.46\% \\ 
$(1,6,\widehat{30},94)$ & 1 & 32 & 49 & 615 &   15.93\% \\ 
$(1,\widehat{6},31,86)$ & 2 & 4 & 0 & 5 &    0.00\% \\ 
$(1,6,31,\widehat{86})$ & 6 & 12 & 0 & 5 &    0.00\% \\ 
$(\widehat{1},6,34,70)$ & 3 & 9 & 24 & 40 &   10.00\% \\ 
$(1,\widehat{6},34,70)$ & 1 & 48 & 35 & 471 &    6.16\% \\ 
$(\widehat{1},6,38,58)$ & 1 & 11 & 32 & 243 &    4.12\% \\ 
$(1,\widehat{6},38,58)$ & 1 & 24 & 59 & 438 &    7.31\% \\ 
$(1,6,38,\widehat{58})$ & 2 & 42 & 0 & 79 &   11.39\% \\ 
$(1,\widehat{6},40,54)$ & 2 & 32 & 0 & 101 &    3.96\% \\ 
$(1,6,40,\widehat{54})$ & 2 & 24 & 0 & 28 &    0.00\% \\ 
$(\widehat{1},6,46,46)$ & 3 & 12 & 101 & 163 &   21.47\% \\ 
$(1,\widehat{6},46,46)$ & 1 & 64 & 69 & 867 &   21.34\% \\ 
$(\widehat{1},7,17,340)$ & 6 & 6 & 0 & 2 &    0.00\% \\ 
$(1,7,\widehat{17},340)$ & 6 & 12 & 0 & 3 &    0.00\% \\ 
$(\widehat{1},7,18,178)$ & 1 & 12 & 31 & 120 &    0.83\% \\ 
$(\widehat{1},7,19,124)$ & 3 & 9 & 26 & 34 &    8.82\% \\ 
$(1,7,\widehat{19},124)$ & 3 & 18 & 4 & 16 &    0.00\% \\ 
$(\widehat{1},7,20,97)$ & 3 & 6 & 4 & 5 &    0.00\% \\ 
$(1,7,\widehat{20},97)$ & 4 & 8 & 0 & 3 &    0.00\% \\ 
$(\widehat{1},7,22,70)$ & 3 & 15 & 121 & 207 &   18.84\% \\ 
$(\widehat{1},7,25,52)$ & 3 & 18 & 50 & 80 &    2.50\% \\ 
$(\widehat{1},7,28,43)$ & 1 & 10 & 60 & 142 &    2.11\% \\ 
$(\widehat{1},7,34,34)$ & 3 & 24 & 123 & 209 &    1.91\% \\ 
$(\widehat{1},8,14,238)$ & 1 & 12 & 56 & 256 &   16.02\% \\ 
$(1,\widehat{8},14,238)$ & 1 & 14 & 12 & 93 &    4.30\% \\ 
$(1,8,\widehat{14},238)$ & 1 & 12 & 21 & 401 &   15.46\% \\ 
$(1,8,14,\widehat{238})$ & 1 & 48 & 0 & 62 &    9.68\% \\ 
$(\widehat{1},8,16,88)$ & 1 & 19 & 64 & 530 &   12.64\% \\ 
$(1,\widehat{8},16,88)$ & 1 & 28 & 43 & 477 &    5.87\% \\ 
$(\widehat{1},8,18,58)$ & 1 & 23 & 726 & 3734 &    6.86\% \\ 
$(1,\widehat{8},18,58)$ & 1 & 32 & 333 & 2234 &    5.73\% \\ 
$(1,8,18,\widehat{58})$ & 1 & 84 & 97 & 1248 &    7.05\% \\ 
$(\widehat{1},8,22,38)$ & 1 & 12 & 12 & 152 &    0.66\% \\ 
$(1,\widehat{8},22,38)$ & 2 & 10 & 0 & 38 &    0.00\% \\ 
$(\widehat{1},8,28,28)$ & 1 & 36 & 1530 & 7738 &   17.19\% \\ 
$(1,\widehat{8},28,28)$ & 1 & 56 & 584 & 4618 &    9.44\% \\ 
$(\widehat{1},9,20,31)$ & 3 & 6 & 8 & 10 &    0.00\% \\ 
$(1,\widehat{9},20,31)$ & 1 & 9 & 1 & 3 &    0.00\% \\ 
$(1,9,\widehat{20},31)$ & 1 & 8 & 2 & 5 &    0.00\% \\ 
$(\widehat{2},10,10,10)$ & 1 & 12 & 28 & 518 &    0.00\% \\ 
$(\widehat{2},2,2,2,2,2)$ & 1 & 48 & 940 & 14131 &    8.90\% \\ 
$(\widehat{2},2,2,3,18)$ & 1 & 24 & 75 & 1029 &   13.70\% \\ 
$(2,2,2,\widehat{3},18)$ & 2 & 30 & 0 & 286 &   11.19\% \\ 
$(\widehat{2},2,2,4,10)$ & 1 & 24 & 185 & 3216 &   13.50\% \\ 
$(2,2,2,\widehat{4},10)$ & 2 & 24 & 0 & 534 &   17.42\% \\ 
$(\widehat{2},2,2,6,6)$ & 1 & 24 & 383 & 9788 &    5.85\% \\ 
$(2,2,2,\widehat{6},6)$ & 1 & 48 & 178 & 3288 &    6.60\% \\ 
$(\widehat{2},2,3,3,8)$ & 1 & 16 & 23 & 443 &    1.35\% \\ 
$(2,2,\widehat{3},3,8)$ & 1 & 24 & 55 & 1112 &    2.34\% \\ 
$(2,2,3,3,\widehat{8})$ & 1 & 32 & 33 & 360 &    0.56\% \\ 
$(2,2,\widehat{4},4,4)$ & 2 & 24 & 0 & 471 &    5.73\% \\ 
$(\widehat{2},3,20,218)$ & 1 & 14 & 0 & 52 &    1.92\% \\ 
$(2,\widehat{3},20,218)$ & 2 & 30 & 0 & 76 &    3.95\% \\ 
$(2,3,\widehat{20},218)$ & 2 & 32 & 0 & 86 &    4.65\% \\ 
$(\widehat{2},3,22,118)$ & 1 & 12 & 24 & 157 &    8.92\% \\ 
$(2,\widehat{3},22,118)$ & 1 & 54 & 10 & 228 &    6.58\% \\ 
$(\widehat{2},3,23,98)$ & 1 & 12 & 32 & 243 &    0.00\% \\ 
$(2,\widehat{3},23,98)$ & 1 & 54 & 30 & 372 &    0.81\% \\ 
$(2,3,\widehat{23},98)$ & 1 & 32 & 24 & 260 &    0.38\% \\ 
$(\widehat{2},3,26,68)$ & 1 & 12 & 0 & 32 &    6.25\% \\ 
$(2,\widehat{3},26,68)$ & 2 & 24 & 0 & 79 &    2.53\% \\ 
$(\widehat{2},3,28,58)$ & 1 & 18 & 169 & 1340 &    3.96\% \\ 
$(2,\widehat{3},28,58)$ & 1 & 78 & 308 & 3150 &   10.63\% \\ 
$(2,3,28,\widehat{58})$ & 1 & 40 & 13 & 137 &    8.03\% \\ 
$(\widehat{2},3,38,38)$ & 1 & 24 & 125 & 1038 &   14.84\% \\ 
$(2,\widehat{3},38,38)$ & 1 & 108 & 166 & 2053 &   10.42\% \\ 
$(\widehat{2},4,11,154)$ & 2 & 4 & 0 & 5 &    0.00\% \\ 
$(2,\widehat{4},11,154)$ & 2 & 8 & 0 & 8 &    0.00\% \\ 
$(\widehat{2},4,13,58)$ & 1 & 8 & 62 & 228 &    4.39\% \\ 
$(2,\widehat{4},13,58)$ & 1 & 12 & 27 & 306 &   12.42\% \\ 
$(2,4,\widehat{13},58)$ & 2 & 30 & 0 & 95 &    2.11\% \\ 
$(2,4,13,\widehat{58})$ & 1 & 32 & 0 & 19 &   15.79\% \\ 
$(\widehat{2},4,14,46)$ & 1 & 12 & 0 & 57 &    8.77\% \\ 
$(2,\widehat{4},14,46)$ & 1 & 12 & 2 & 29 &   10.34\% \\ 
$(2,4,\widehat{14},46)$ & 2 & 12 & 0 & 68 &    0.00\% \\ 
$(2,\widehat{4},16,34)$ & 1 & 8 & 8 & 248 &    9.27\% \\ 
$(\widehat{2},4,18,28)$ & 1 & 12 & 18 & 161 &    0.00\% \\ 
$(2,\widehat{4},18,28)$ & 1 & 10 & 0 & 79 &    2.53\% \\ 
$(\widehat{2},4,22,22)$ & 1 & 12 & 0 & 61 &    0.00\% \\ 
$(2,\widehat{4},22,22)$ & 1 & 8 & 2 & 24 &    0.00\% \\ 
$(\widehat{2},5,10,40)$ & 3 & 12 & 4 & 19 &    0.00\% \\ 
$(\widehat{2},5,12,26)$ & 1 & 12 & 22 & 141 &    0.71\% \\ 
$(2,5,\widehat{12},26)$ & 1 & 5 & 10 & 40 &    0.00\% \\ 
$(2,5,12,\widehat{26})$ & 1 & 15 & 4 & 20 &   10.00\% \\ 
$(\widehat{2},5,8,138)$ & 1 & 6 & 2 & 69 &    8.70\% \\ 
$(2,\widehat{5},8,138)$ & 1 & 6 & 1 & 35 &    2.86\% \\ 
$(2,5,\widehat{8},138)$ & 1 & 6 & 4 & 10 &    0.00\% \\ 
$(\widehat{2},6,10,22)$ & 1 & 15 & 19 & 426 &    2.11\% \\ 
$(2,\widehat{6},10,22)$ & 1 & 12 & 24 & 248 &    0.40\% \\ 
$(\widehat{2},6,14,14)$ & 1 & 20 & 0 & 138 &    2.17\% \\ 
$(2,\widehat{6},14,14)$ & 1 & 32 & 32 & 224 &    3.12\% \\ 
$(2,6,\widehat{14},14)$ & 1 & 18 & 67 & 350 &    5.71\% \\ 
$(2,6,14,\widehat{14})$ & 1 & 18 & 55 & 348 &    2.87\% \\ 
$(\widehat{2},6,7,70)$ & 1 & 6 & 0 & 20 &   10.00\% \\ 
$(2,\widehat{6},7,70)$ & 2 & 2 & 0 & 11 &    9.09\% \\ 
$(\widehat{2},6,8,38)$ & 1 & 16 & 48 & 387 &    2.58\% \\ 
$(2,\widehat{6},8,38)$ & 1 & 22 & 18 & 320 &    3.12\% \\ 
$(2,6,\widehat{8},38)$ & 1 & 18 & 1 & 140 &    0.00\% \\ 
$(\widehat{2},7,10,16)$ & 1 & 10 & 14 & 114 &    1.75\% \\ 
$(\widehat{2},7,7,34)$ & 1 & 6 & 2 & 8 &    0.00\% \\ 
$(\widehat{2},8,10,13)$ & 1 & 8 & 2 & 61 &    4.92\% \\ 
$(2,\widehat{8},10,13)$ & 4 & 8 & 0 & 6 &    0.00\% \\ 
$(2,8,10,\widehat{13})$ & 2 & 22 & 0 & 28 &    0.00\% \\ 
$(\widehat{2},8,8,18)$ & 1 & 24 & 311 & 2310 &    1.90\% \\ 
$(2,\widehat{8},8,18)$ & 1 & 24 & 395 & 3115 &    1.28\% \\ 
$(\widehat{3},3,10,58)$ & 1 & 30 & 80 & 486 &    0.00\% \\ 
$(3,3,10,\widehat{58})$ & 1 & 16 & 5 & 49 &    0.00\% \\ 
$(\widehat{3},3,12,33)$ & 1 & 16 & 2 & 89 &    0.00\% \\ 
$(3,3,\widehat{12},33)$ & 2 & 12 & 0 & 9 &    0.00\% \\ 
$(\widehat{3},3,13,28)$ & 1 & 54 & 1034 & 6388 &    5.42\% \\ 
$(3,3,\widehat{13},28)$ & 1 & 48 & 145 & 999 &    0.20\% \\ 
$(\widehat{3},3,18,18)$ & 1 & 68 & 1802 & 12434 &    1.17\% \\ 
$(\widehat{3},3,3,3,3)$ & 1 & 50 & 3189 & 25714 &   20.91\% \\ 
$(\widehat{3},3,9,108)$ & 1 & 20 & 3 & 92 &    1.09\% \\ 
$(3,3,\widehat{9},108)$ & 2 & 12 & 0 & 12 &    0.00\% \\ 
$(\widehat{3},4,10,18)$ & 2 & 18 & 0 & 76 &    0.00\% \\ 
$(3,\widehat{4},10,18)$ & 2 & 8 & 0 & 20 &    0.00\% \\ 
$(\widehat{3},4,13,13)$ & 1 & 48 & 11 & 138 &    0.00\% \\ 
$(3,\widehat{4},13,13)$ & 2 & 12 & 0 & 83 &    0.00\% \\ 
$(3,4,\widehat{13},13)$ & 1 & 30 & 16 & 79 &    0.00\% \\ 
$(\widehat{3},4,6,118)$ & 1 & 18 & 0 & 203 &    4.93\% \\ 
$(3,\widehat{4},6,118)$ & 2 & 8 & 0 & 48 &    0.00\% \\ 
$(3,4,\widehat{6},118)$ & 1 & 2 & 0 & 84 &    1.19\% \\ 
$(\widehat{3},4,7,43)$ & 1 & 24 & 4 & 16 &    0.00\% \\ 
$(3,\widehat{4},7,43)$ & 2 & 4 & 0 & 5 &    0.00\% \\ 
$(\widehat{3},4,8,28)$ & 1 & 48 & 340 & 2638 &    4.66\% \\ 
$(3,\widehat{4},8,28)$ & 1 & 20 & 71 & 1116 &    1.43\% \\ 
$(3,4,\widehat{8},28)$ & 1 & 32 & 86 & 644 &    1.24\% \\ 
$(\widehat{3},5,5,68)$ & 1 & 24 & 0 & 27 &    0.00\% \\ 
$(3,\widehat{5},5,68)$ & 1 & 8 & 9 & 13 &    0.00\% \\ 
$(3,5,\widehat{5},68)$ & 1 & 6 & 9 & 17 &    0.00\% \\ 
$(\widehat{3},6,6,18)$ & 2 & 28 & 0 & 158 &    0.63\% \\ 
$(3,\widehat{6},6,18)$ & 1 & 8 & 18 & 99 &    0.00\% \\ 
$(\widehat{3},8,8,8)$ & 1 & 72 & 3046 & 23592 &    4.80\% \\ 
$(3,\widehat{8},8,8)$ & 1 & 32 & 3137 & 21945 &    3.20\% \\ 
$(\widehat{4},4,10,10)$ & 1 & 20 & 45 & 420 &    2.86\% \\ 
$(\widehat{4},4,5,40)$ & 1 & 12 & 0 & 64 &    3.12\% \\ 
$(\widehat{4},4,6,22)$ & 1 & 14 & 4 & 260 &    5.38\% \\ 
$(4,4,\widehat{6},22)$ & 4 & 8 & 0 & 26 &    3.85\% \\ 
$(\widehat{4},4,7,16)$ & 1 & 18 & 104 & 491 &    3.87\% \\ 
$(\widehat{4},4,8,13)$ & 1 & 12 & 5 & 166 &    1.81\% \\ 
$(4,4,\widehat{8},13)$ & 4 & 16 & 0 & 23 &    0.00\% \\ 
$(4,4,8,\widehat{13})$ & 2 & 28 & 0 & 25 &    0.00\% \\ 
$(\widehat{4},5,5,19)$ & 4 & 8 & 0 & 2 &    0.00\% \\ 
$(4,\widehat{5},5,19)$ & 1 & 8 & 1 & 2 &    0.00\% \\ 
$(4,5,\widehat{5},19)$ & 3 & 3 & 1 & 1 &    0.00\% \\ 
$(\widehat{4},6,6,10)$ & 2 & 2 & 0 & 2 &    0.00\% \\ 
$(4,\widehat{6},6,10)$ & 4 & 8 & 0 & 9 &    0.00\% \\ 
$(\widehat{4},7,7,7)$ & 2 & 6 & 0 & 14 &    0.00\% \\ 
$(\widehat{5},5,5,12)$ & 1 & 8 & 13 & 82 &    0.00\% \\ 
$(5,\widehat{5},5,12)$ & 3 & 6 & 29 & 40 &    0.00\% \\ 
$(5,5,5,\widehat{12})$ & 1 & 8 & 6 & 22 &    0.00\% \\ 
$(\widehat{6},6,6,6)$ & 1 & 16 & 174 & 1869 &    0.27\% \\ 
$(1,10,\widetilde{11},154)$ & 2 & 12 & 0 & 23 &    0.00\% \\ 
$(1,10,\widetilde{14},46)$ & 1 & 14 & 83 & 599 &   11.85\% \\ 
$(1,1,1,1,\widetilde{5},40)$ & 6 & 12 & 0 & 5 &   20.00\% \\ 
$(1,1,2,\widetilde{14},46)$ & 1 & 12 & 20 & 195 &    0.00\% \\ 
$(1,1,4,\widetilde{5},40)$ & 6 & 12 & 0 & 13 &    0.00\% \\ 
$(1,\widetilde{5},42,922)$ & 3 & 24 & 4 & 18 &    0.00\% \\ 
$(1,\widetilde{5},43,628)$ & 1 & 24 & 6 & 24 &    4.17\% \\ 
$(1,\widetilde{5},46,334)$ & 3 & 24 & 27 & 62 &    1.61\% \\ 
$(1,\widetilde{5},47,292)$ & 3 & 24 & 2 & 7 &    0.00\% \\ 
$(1,\widetilde{5},52,187)$ & 6 & 12 & 0 & 7 &   28.57\% \\ 
$(1,\widetilde{5},54,166)$ & 3 & 30 & 27 & 61 &    0.00\% \\ 
$(1,\widetilde{5},58,138)$ & 1 & 12 & 10 & 33 &    0.00\% \\ 
$(1,\widetilde{5},61,124)$ & 3 & 36 & 10 & 19 &    0.00\% \\ 
$(1,\widetilde{5},68,103)$ & 1 & 18 & 0 & 26 &   11.54\% \\ 
$(1,\widetilde{5},82,82)$ & 3 & 48 & 39 & 74 &    1.35\% \\ 
$(1,7,\widetilde{17},340)$ & 2 & 8 & 0 & 7 &    0.00\% \\ 
$(1,8,\widetilde{14},238)$ & 1 & 12 & 16 & 428 &   23.13\% \\ 
$(2,4,\widetilde{14},46)$ & 2 & 24 & 0 & 72 &    1.39\% \\ 
$(2,\widetilde{5},8,138)$ & 1 & 8 & 2 & 40 &    0.00\%   
\end{longtable}
\end{center}

\section{Results on Vector-like Particles}

As already emphasized before, we regard the chiral data (the number of families and the
absence of chiral exotics) as the main area of study for these methods. However, we did also 
collect non-chiral data such as the number of mirrors, singlets and fractionally charged vector-like matter. 
The rationale for doing that is to determine how many distinct points in a given moduli space we obtain. 
Ideally, one could do that by comparing MIPFs. But this is hard to do. In principle any distinct simple
current subgroup and/or any different choice of discrete torsion parameters defines a different MIPF, but
there are usually large degeneracies, which are not all understood, and are difficult to ``mod out" even if they are. 
So the simpler question we consider is: how many distinct massless spectra do we get (with ``distinct" defined as in section 4).
Obviously two identical massless spectra may still belong to different MIPFs, but when we made finer distinctions 
(as discussed in the next section) we found that in most cases a given spectrum yielded only one distinct refined spectrum.
Hence it seems that our results give a pretty good estimate for the true number of distinct MIPFs.

Having gathered all this data we can also use them for a different purpose, namely to plot distributions of 
non-chiral quantities. 
The most important lesson that may be drawn from this procedure is how easy it would be to find exact 
RCFT spectra with certain desired
properties, such as absence of certain vector-like exotics. Although this problem should really be
addressed in a continuous  geometric approach,  as we have emphasized before, at least the existence 
of an exact RCFT example demonstrates that there are
no fundamental obstacles for finding such solutions.

\begin{figure}[P]
\begin{center}
\includegraphics[width=13cm]{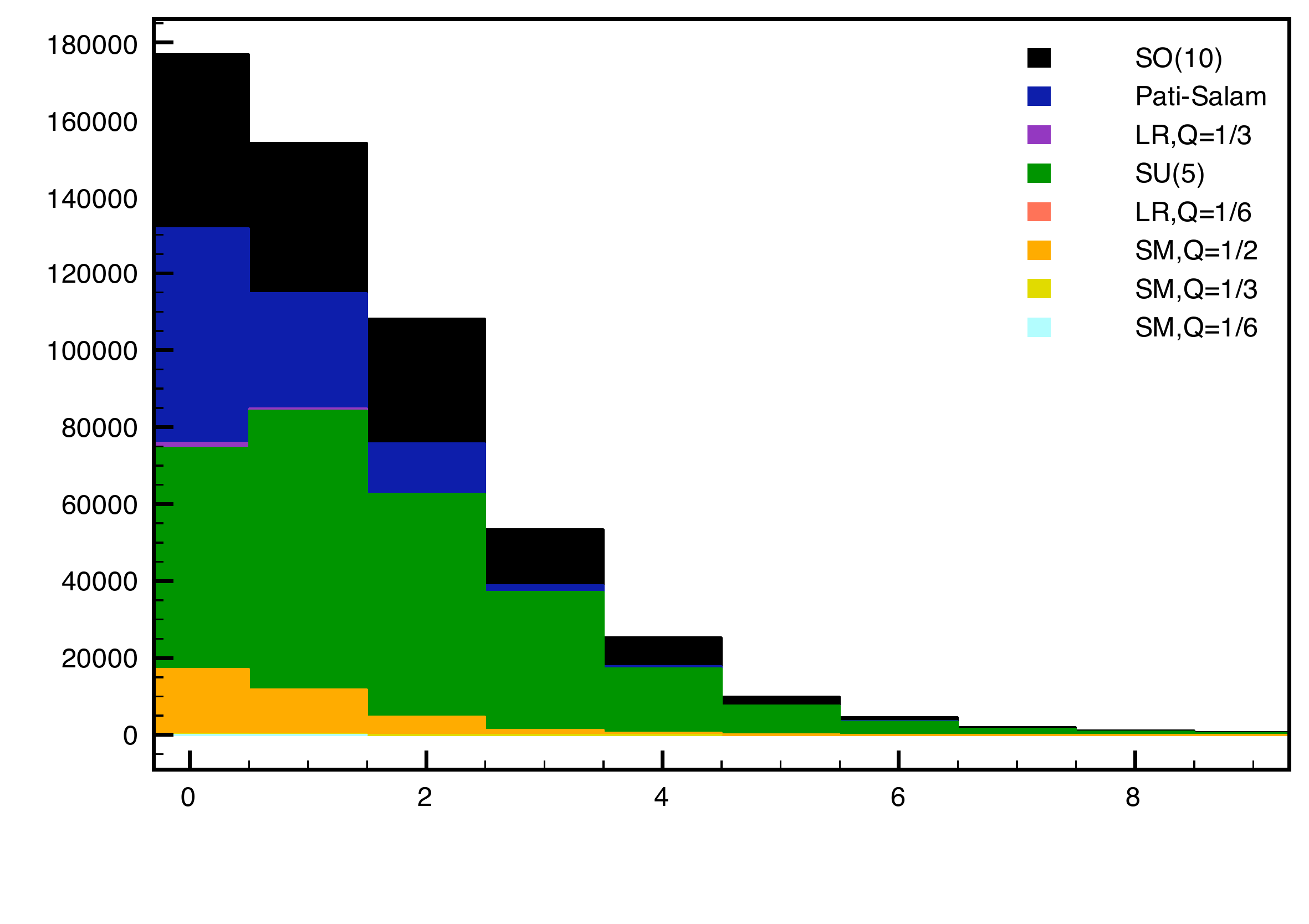}
\caption{Q-Mirror distribution.}
\label{QMirror}
\end{center}
\end{figure}

\begin{figure}[P]
\begin{center}
\includegraphics[width=13cm]{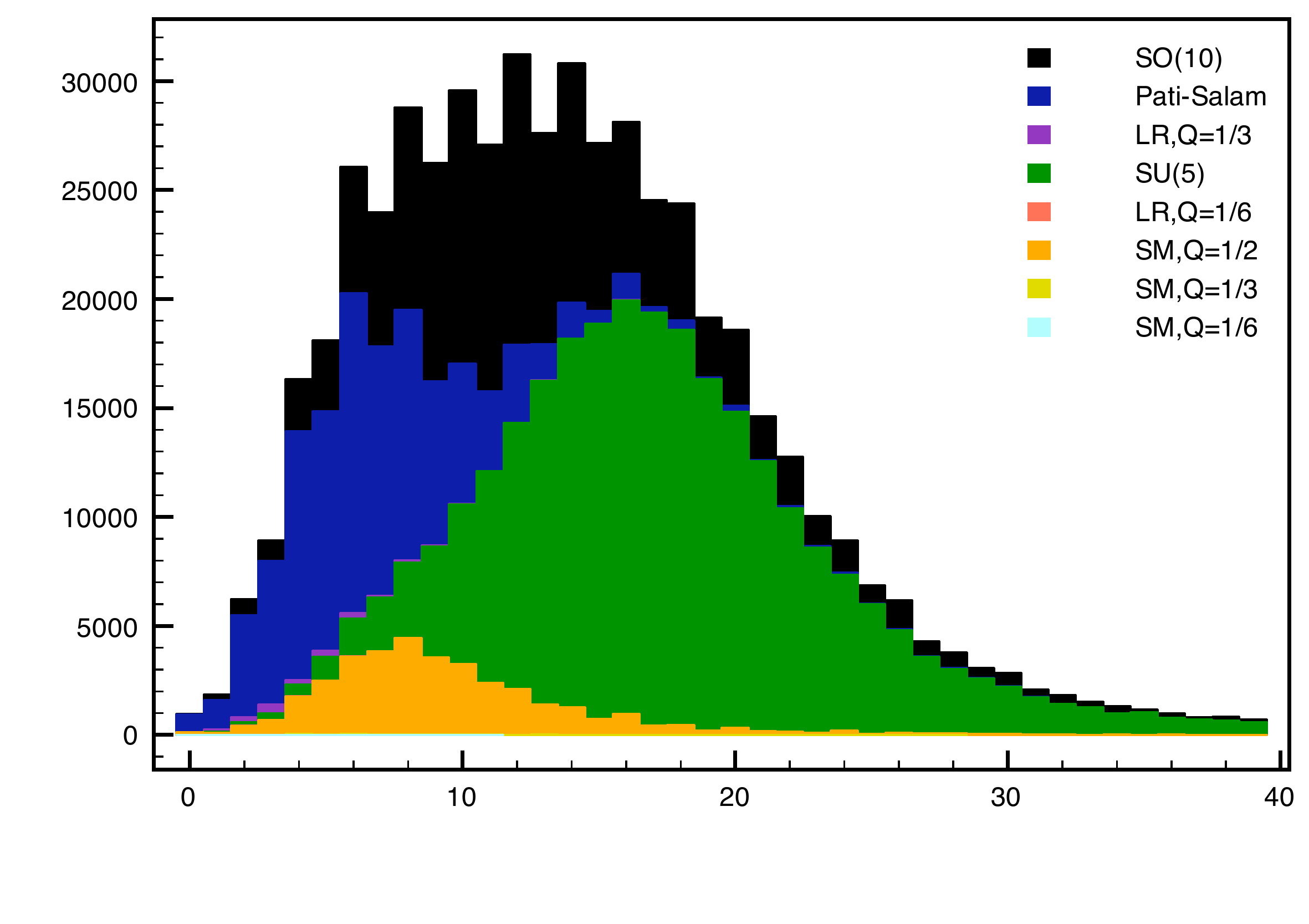}
\caption{L-Mirror distribution.}
\label{LMirror}
\end{center}
\end{figure}

In all of these plots we display on the vertical axis the number of distinct MIPFs, counting only models 
without chiral exotics, and
with at least one family. We treat mirror symmetric MIPFs as distinct. Furthermore, all plots are 
stacked histograms: the contributions
of the 8 different group types are stacked on top of each other for better visibility.
In Figs. \ref{QMirror} and \ref{LMirror} we show respectively the distribution of Q and L-type mirrors. 
It is noteworthy
that both distributions have shifted towards zero in comparison to the standard Gepner model 
case \cite{GatoRivera:2010gv}. For Pati-Salami models,
the Q mirror distribution peaks at  4 for standard Gepner models, and at 0 for lifted ones. 
For SM, Q=$\frac12$ models the Q mirrors peak at 3 (standard) and 0 (lifted). 
So in both cases the peak is at the MSSM value. For L mirror pair the MSSM value is one (the Higgs pair $H_1, H_2$);
the peaks do indeed shift in that direction, from a value of around 20 to values of around 10. 

\begin{figure}[P]
\begin{center}
\includegraphics[width=13cm]{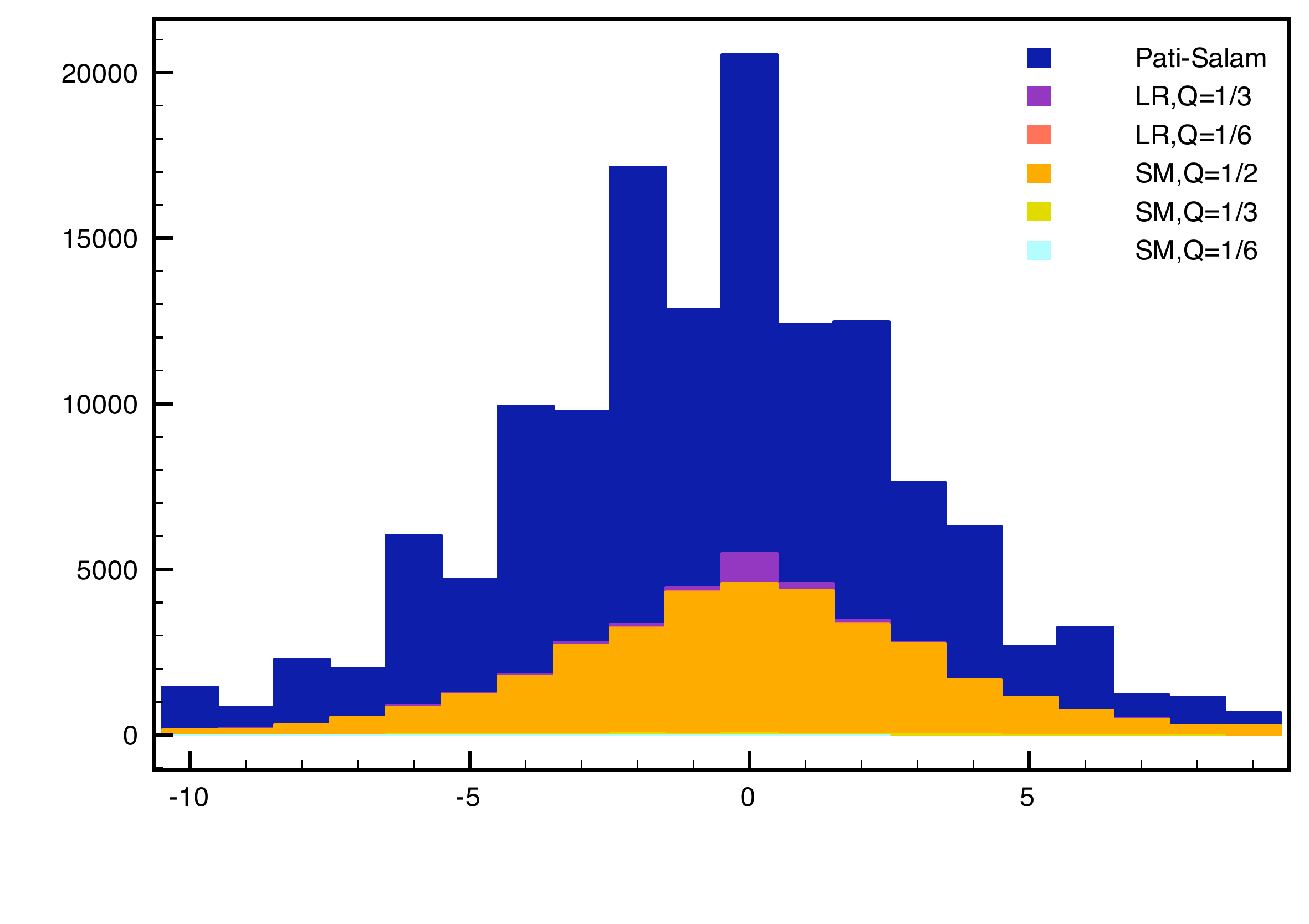}
\caption{Difference between the D and L distributions.}
\label{DLDif}
\end{center}
\end{figure}

\begin{figure}[P]
\begin{center}
\includegraphics[width=13cm]{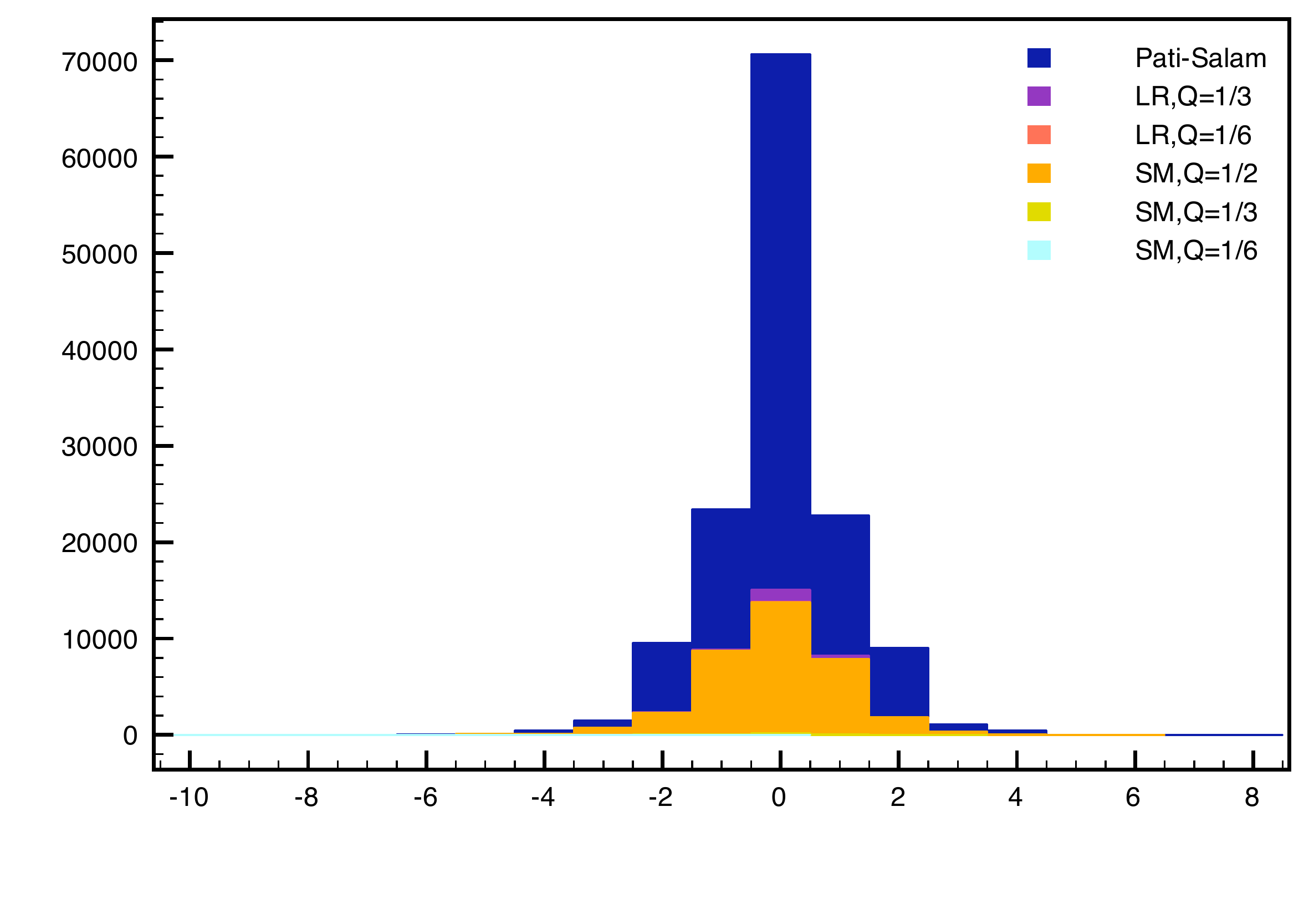}
\caption{Difference between the Q and E distributions.}
\label{QEDif}
\end{center}
\end{figure}

The analogous plots for U and E mirrors are practically indistinguishable from those of Q mirrors, and the D-mirror plot
is nearly identical to the L-mirror plot. This is because in $SU(5)$ GUTs, Q, U and E all come from the representation $(10)$, while
D, L both come from the $(\bar 5)$. Hence for the dominant contribution to the plots these distributions are identical.
But even for the broken GUTs these quantities are not all unrelated: the spin-2 current $(3,1,10,0)$ of 
$SU(3) \times SU(2) \times U(1)_{30} \times U(1)_{20}$, that is always present in the half-integer fractional charge models, relates
U and E. Since the half-integer fractional charge models (Pati-Salam and SM, Q=$\frac12$) dominate the statistics for the non-GUT
models, there are only two differences that can still be non-trivial: D-L and Q-E. We show the distributions for these differences in 
Figs. \ref{DLDif} and \ref{QEDif} respectively. The main conclusion is that even in broken GUT models there is a tendency for
mirror particles to be in complete GUT multiplets. This has the interesting consequence that mirror particles tend not
to affect the convergence of couplings. The effect is somewhat weaker for members of the $(\bar 5)$, but it is still no entirely
unreasonable to assume that the effect of mirrors on coupling unification could be small. 

\begin{figure}[P]
\begin{center}
\includegraphics[width=12cm]{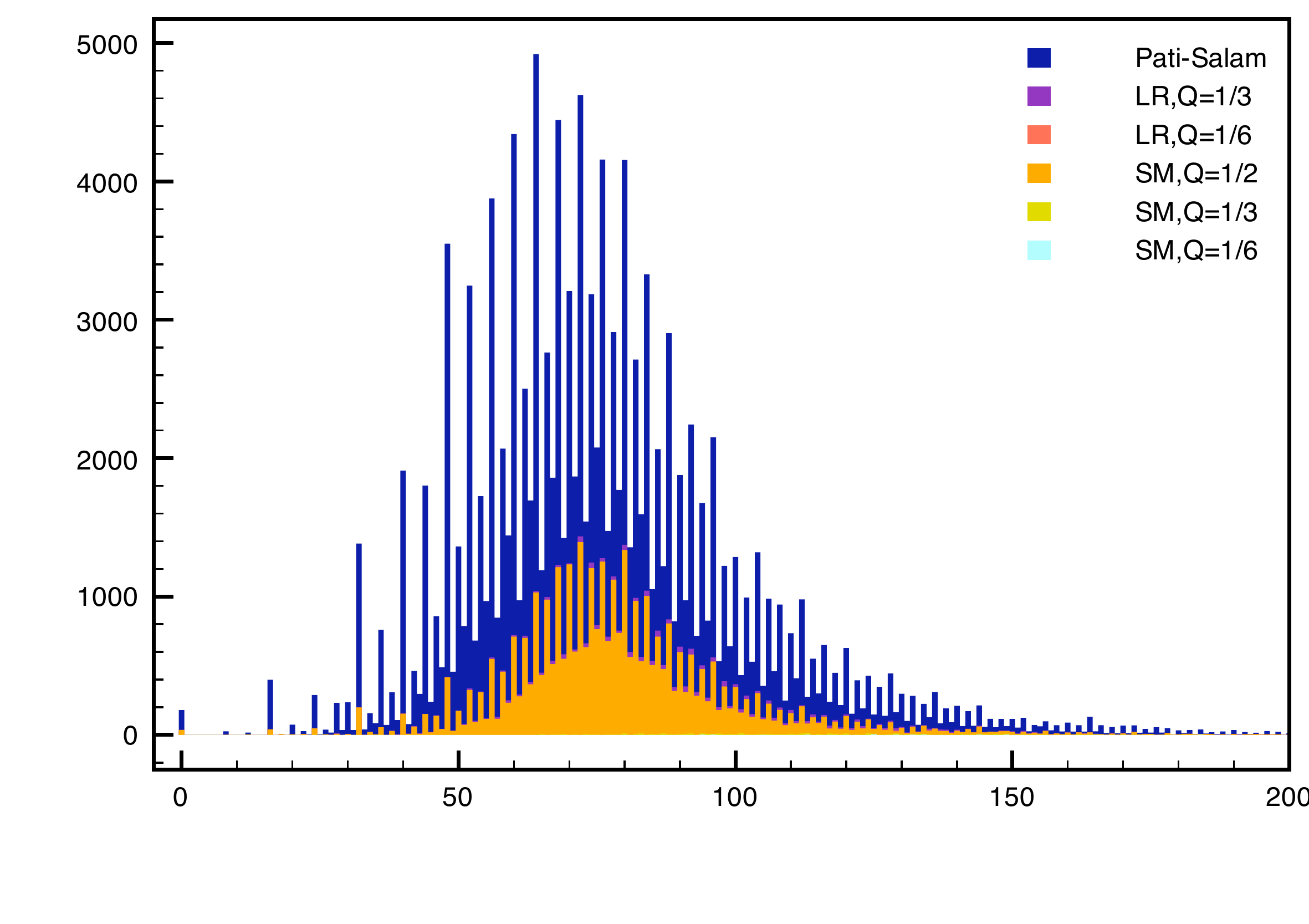}
\caption{ The number of MIPFs with a certain number of vector-like pairs of fractionally charged particles.}
\label{Frac}
\end{center}
\end{figure}

In Fig. \ref{Frac} we show the distribution of fractionally charged particles. What is counted here along the x-axis are half the number of
$SU(3)\times SU(2) \times U(1)$ representations. The SM-dimensions (color triplets, weak doublets) 
of those representations are not taken into account, but
the dimensions with respect to any additional gauge factor are counted as multiplicities. Most vector-like fractionally charged
exotics come in pairs, but the half-integer charged representation $(1,2,0)$ is real by itself (this is the only 
available $SU(3)\times SU(2) \times U(1)$ representation, apart from the singlet, with that property). 
However, it too must appear with even multiplicities,
since otherwise there would be an $SU(2)$ global anomaly (note that this anomaly cancel within each chiral family, and among the
two members of any other vector-like pair). For this reason we have divided the number of fractionals by two, so that essentially
what is shown is the number of vector-like pairs. There is an interesting structure visible in these plots.
There are considerably more SM, Q=$\frac12$ models where the number of vector-like pairs is even. For the Pati-Salam models
this goes even further, with a successive enhancement if the number of pairs is even, a multiple of four and a multiple of eight. The
origin of this phenomenon is not clear.

The singlet distribution is shown in Fig. \ref{Singlet}. The positions of the peaks of these distributions are not very different from those
of the standard Gepner model case \cite{GatoRivera:2010gv}, but the distributions are much broader.

One might have hoped that weight lifting, since it increases more conformal weights than it decreases, would lower
the number of singlets  (as indeed it seems to do for the mirrors). However, one should keep in mind that not only
the conformal weight changes, but that also the ground state dimensions in general increase, because the ground states
are often in non-trivial representations of the extra non-abelian gauge groups of the lifted CFT. When counting states we did
take these dimensions into account. Therefore the actual number of singlet particle representations is considerably smaller. This
effect is more important for standard model singlets than for mirrors, because the latter get a contribution to their conformal
weight from the standard model representation they carry, so that there is less ``space" left for a non-trivial representation 
of a non-abelian factor: the total conformal weight has to add up to 1.

 \section{Non-abelian Group Representations}
 
\begin{figure}[P]
\begin{center}
\includegraphics[width=13cm]{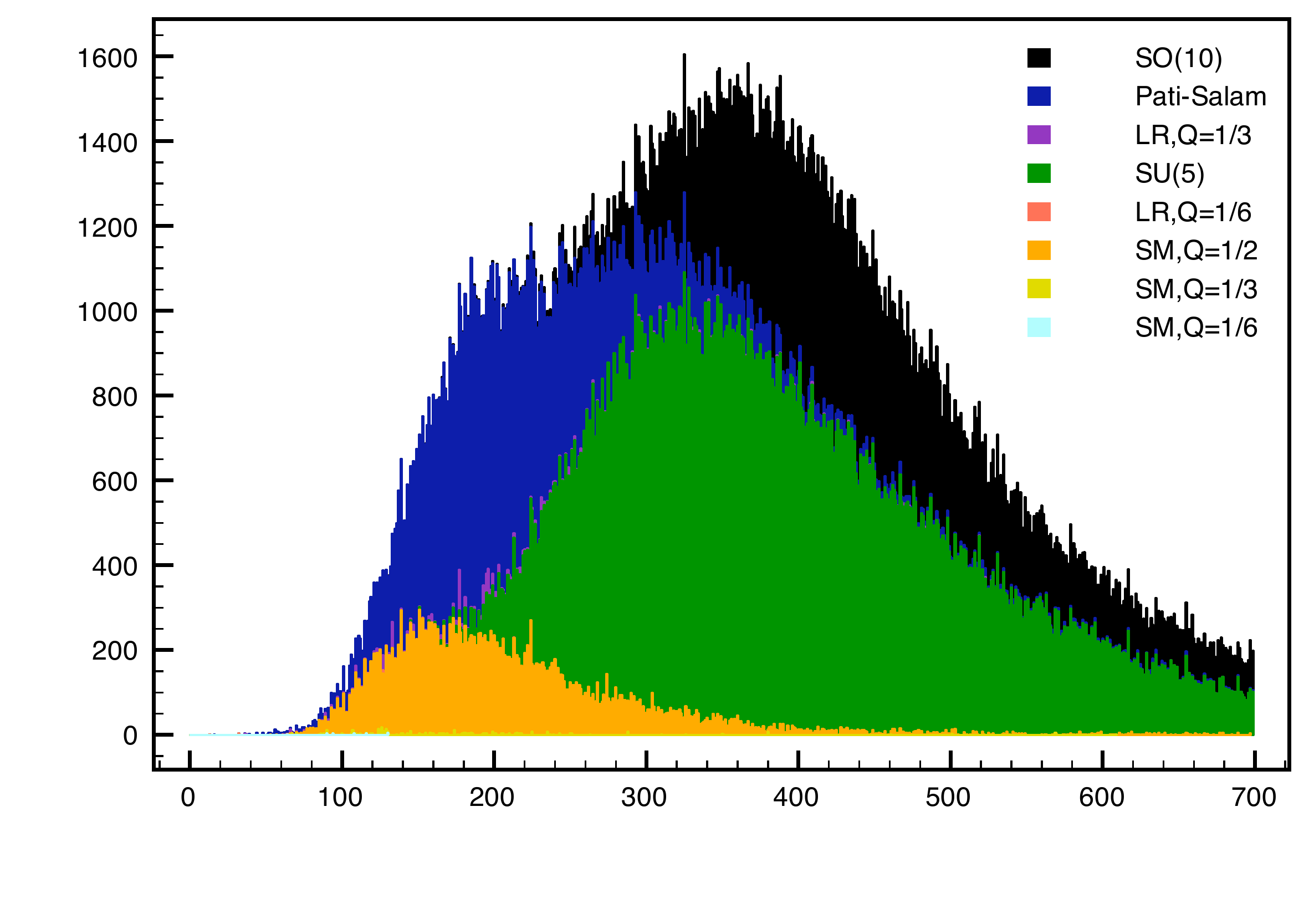}
\caption{Singlet distribution.}
\label{Singlet}
\end{center}
\end{figure}

\begin{figure}[P]
\begin{center}
\includegraphics[width=13cm]{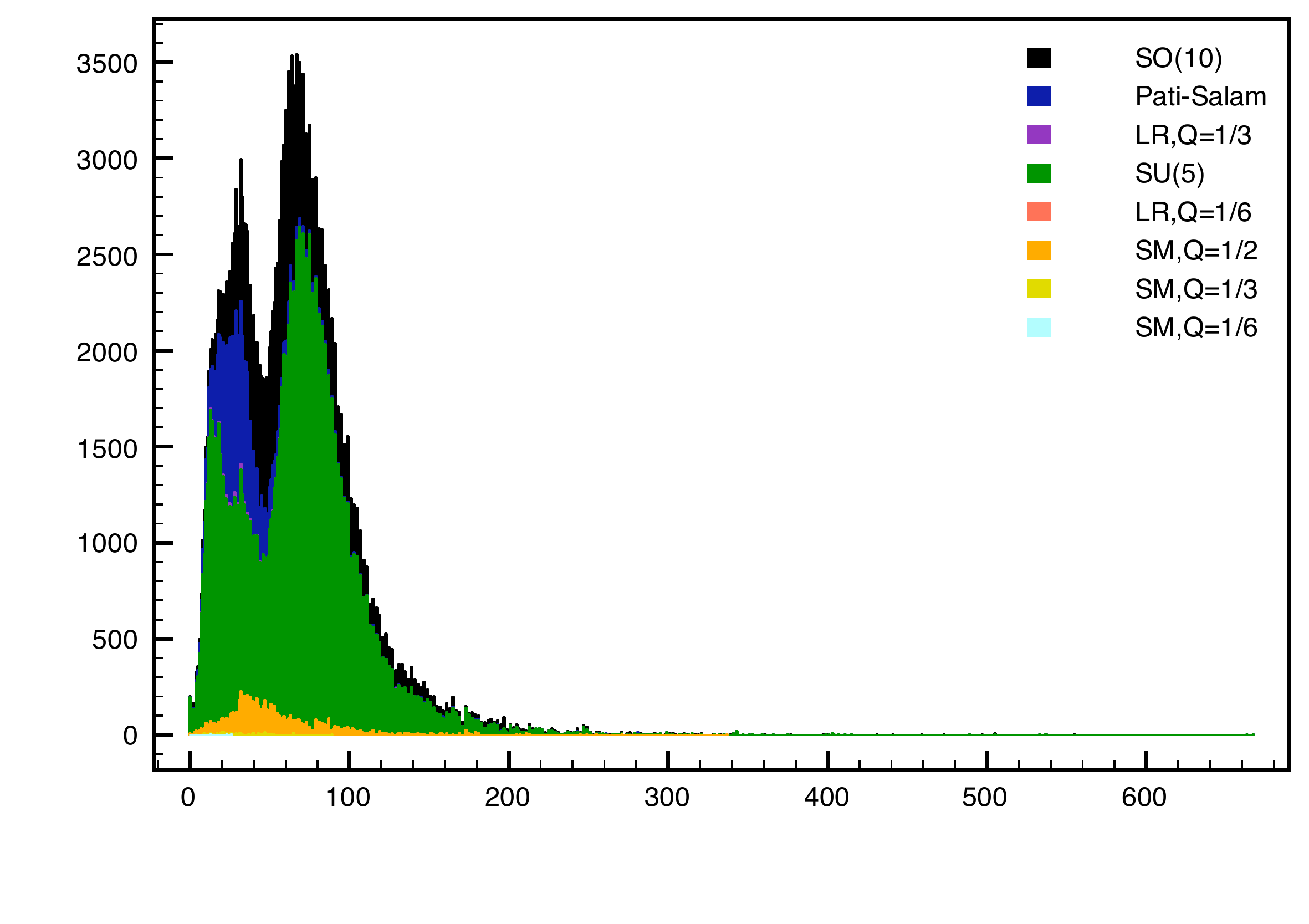}
\caption{Abelian singlet distribution: singlets that do not couple to non-abelian hidden sector groups.}
\label{AbelianSinglet}
\end{center}
\end{figure}

In orientifold models, matter that is charged under additional gauge group factors but not under the standard model
is usually not counted as a singlet, but as hidden sector matter. 
One can perform a similar count here. In Fig. \ref{AbelianSinglet} we plot the distribution of standard model matter singlets 
(all singlets except the
dilaton multiplet)
that are in the trivial representation of the non-abelian factors of the ``hidden sector". Here we are only considering non-abelian
groups originating from the lifts. In general there will also be a few cases where the chiral algebra of the internal sector of the theory
(that is, the part of the CFT outside $SO(10)$) is extended to a non-abelian group, but these extensions were not taken into account.
As expected, the distributions move towards zero in comparison to Fig. \ref{Singlet}, and there are even some cases where 
zero is reached, so that {\it all} standard model matter singlets are in a non-trivial representation of some non-abelian hidden sector group. 

The latter are examples of chiral heterotic spectra where the only gauge singlets are the dilaton and the axion, 
$B_{\mu\nu}$ (and of course
the graviton).
 However, this does not mean that there are no moduli,  apart from the dilaton multiplet. One can observe
the same phenomenon in enhanced symmetry points in the moduli space of Narain compactifications. In that case the moduli
are in non-trivial representations of non-abelian groups, and giving them a v.e.v. breaks the gauge symmetry. It is likely that the same
happens here and that these models correspond to enhanced symmetry points in a larger moduli space. On the other hand, if most
or all of the moduli are in non-trivial non-abelian gauge group representations, this opens the way to dynamical mechanisms to lift them
when these gauge interactions become strong. Note that even if standard model singlets are also singlets with respect to all {\it non-abelian}
groups originating from the lift, they will usually still be charged under {\it abelian} factors in the internal sector. Indeed, in standard Gepner
models most singlets (and in particular most moduli) are charged. In this case we know from comparison with the geometric case
that moving in moduli space will break some or all of the $U(1)$'s.  For this reason we ignored such $U(1)$ charges when counting singlets.

Analogously, one may ask if the fractionally charged matter could be mostly in non-abelian gauge representations. Indeed, one
way around the problem of the fractionally charged particles would be that all of them are confined by some additional gauge force
into integer charged particles. Although this looks like an ugly and far-fetched solution (and probably is) to a problem that is beautifully
solved by field-theoretic $SU(5)$ GUTs, remarkably the first example of heterotic weight lifting we worked out in detail in 
\cite{GatoRivera:2009yt} had precisely that feature. In that example, occurring in the tensor product $(3,\widehat{8},8,8)$, 
the gauge group contains a factor $A_{1,8}$ ($SU(2)$ level 8), and
in the massless spectrum all half-integer charged particles turned out to be in half-integer spin representations of this $SU(2)$, whereas
all integer-charged particles are in integer spin representations (this was only checked for the massless spectrum). 
 
\begin{figure}[P]
\begin{center}
\includegraphics[width=18cm]{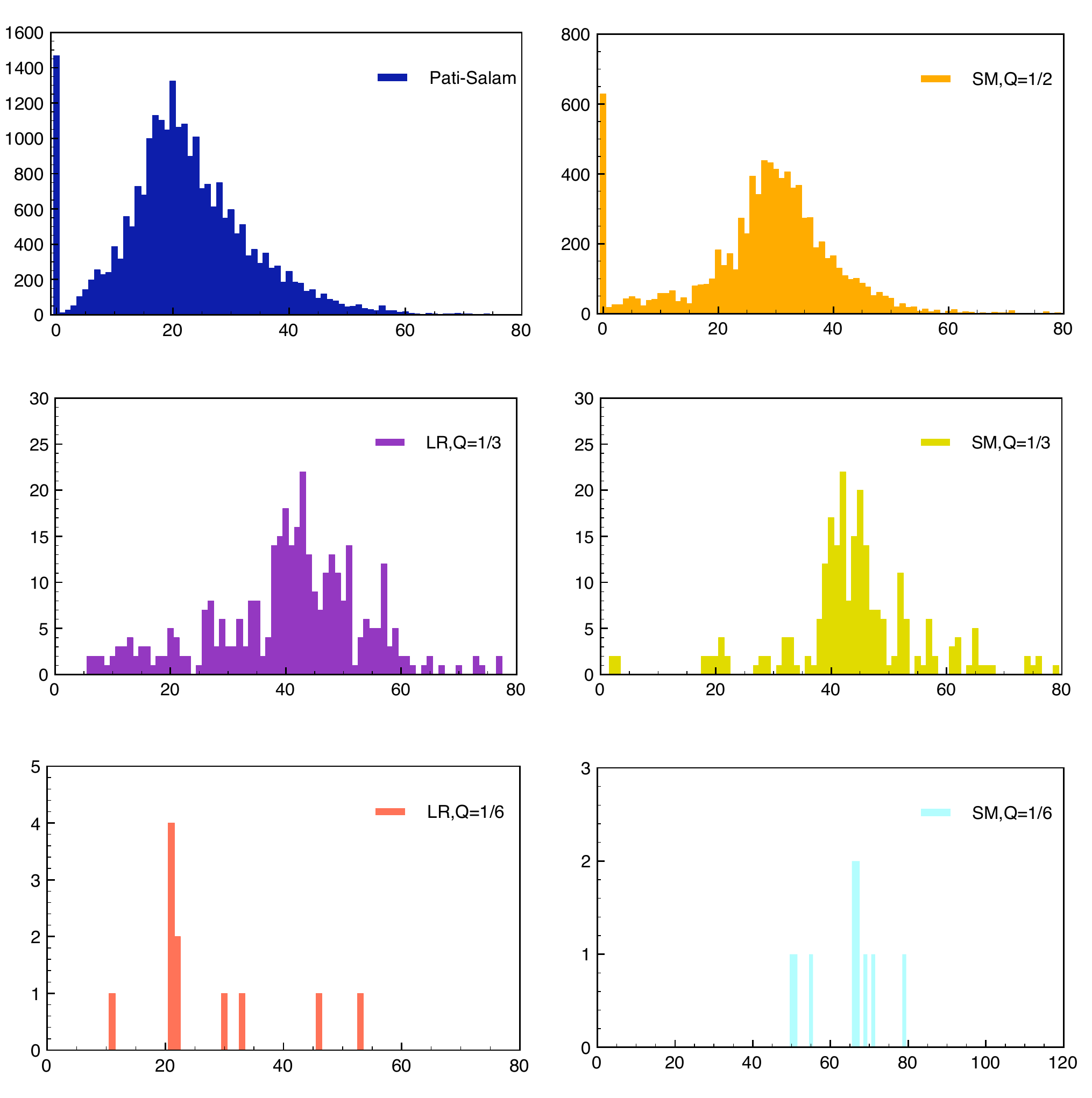}
\caption{Abelian fractional charge distributions.}
\label{AbFrac}
\end{center}
\end{figure}

 In the present work we have merely checked if in general the fractionally charged matter is in non-trivial 
 representations of non-abelian
 groups. The answer turns out to be mostly negative (even if it had been positive, this would still not guarantee 
 that the gauge group confines them).
 The result is shown in Fig. \ref{AbFrac}. These distributions show, for all six types of broken GUTs, the number of MIPFs (y-axis)
 with a certain number of vector-like pairs of fractionally charged particles that are singlets with respect to the extra non-abelian
 factors. Clearly in the majority of cases there are at least some fractionally charged particles that are singlets of all 
 non-abelian hidden sector groups, and hence cannot be
 confined. Interestingly, for the half-integer charge models there are large peaks  at zero, which are completely absent for the
 third- and sixth integer charge models (of which there are very few; please note that the scales of these plots are very different).
 For the models contributing to these peaks all fractionally charged particles couple to non-abelian groups. 
 However, as already remarked above, this still does not mean
 that they are confined, 
 and anyhow this is only a small part of the total surface area.

\begin{figure}[P]
\begin{center}
\includegraphics[width=13cm]{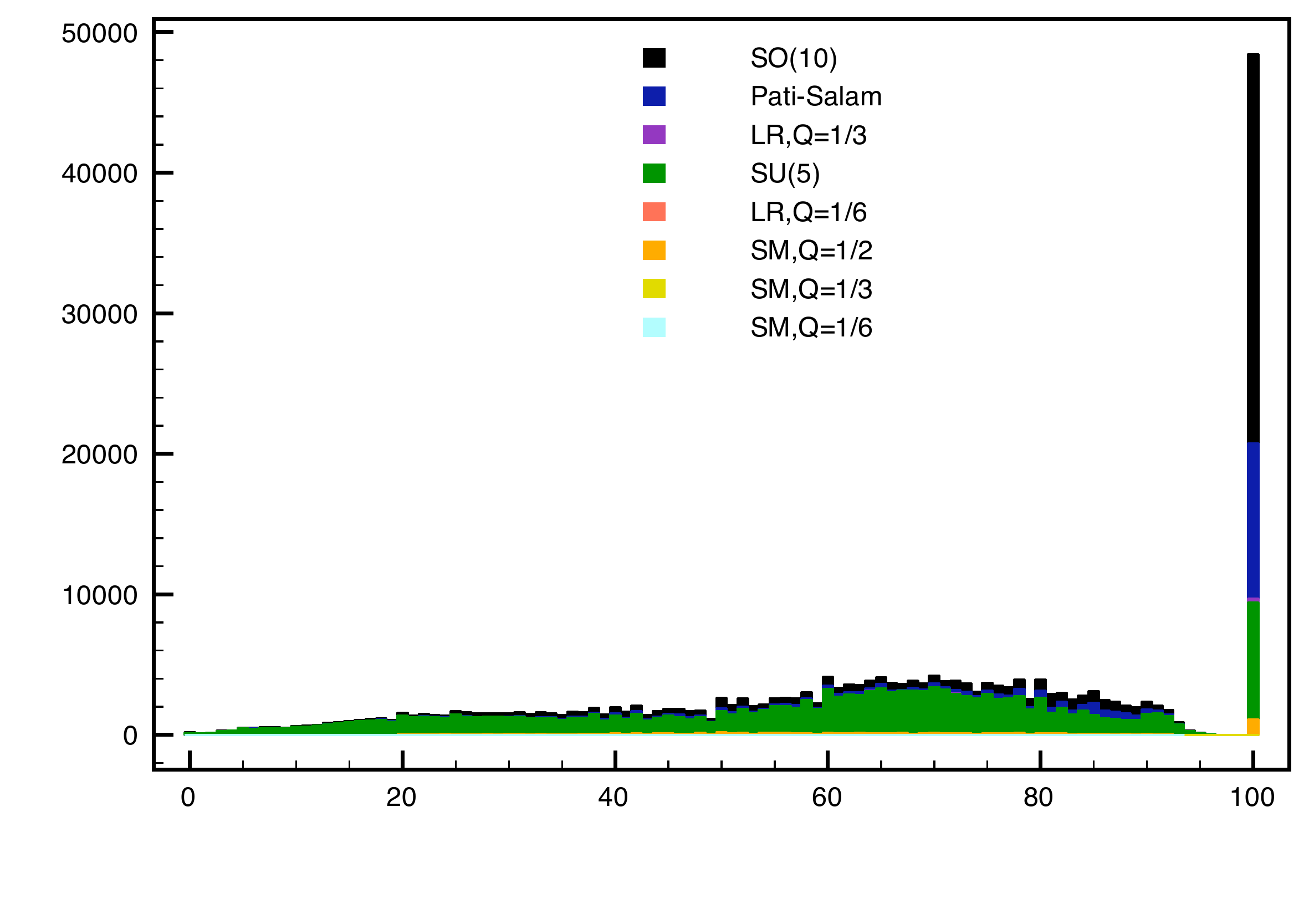}
\caption{Percentage of all matter that is abelian with respect to the extra gauge factors.}
\label{AbFracMat}
\end{center}
\end{figure}

 Finally, one may ask if the standard model families couple to the additional non-abelian groups. If they do, and if those non-abelian
 groups remain unbroken, this will almost certainly give rise to serious phenomenological problems. 
 For example, in the aforementioned model
 presented in \cite{GatoRivera:2009yt} some of the standard model matter is in triplets of $SU(2)_8$, in such a way that
 some Yukawa couplings are forbidden. Here we have simply checked which fraction of the total amount of matter (families and
 mirror pairs) is fully abelian in the extra gauge groups. The result is shown in Fig. \ref{AbFracMat}. There is a huge peak at $100\%$,
 representing spectra for which all matter is in the singlet representation of the extra non-abelian groups. Many of the other
 spectra are not yet ruled out though: there will also be examples where only the mirror fermions couple to the extra gauge groups.
 This is not a problem and might even be a benefit. 
 
 However, we have not investigated such more detailed questions. The most
 important issue is whether the extra non-abelian groups introduced by heterotic weight lifting give rise to such serious obstacles
 that the whole class must be rejected. This seems not to be the case.

\section{Conclusions}

We have studied a new region of the heterotic landscape that is a bit  more genuinely $(0,2)$ than standard Gepner models. This
immediately solves an annoying problem of the latter: the difficulty to get three families. Other classes we have studied (and which will
be published in future work) seem
to confirm this. 
If indeed the distributions we get for these
models are typical for heterotic strings, then the observed number of families is not an issue of concern. 
The family distribution is fairly smooth around three families and does not have the sharp dip observed in 
orientifold models. Nevertheless, from
these two different results in two accessible corners of the string landscape no conclusion can be drawn about such distributions
on the full landscape.

What does remain an issue of concern is the occurrence of fractional charges. Unless in generic heterotic vacua 
vector-like particles are 
lifted to scales far above the weak scale, we have not found any convincing reason for their absence or extremely 
small abundance in our environment. In the
class we studied here, as well as those studied in \cite{GatoRivera:2010gv} and \cite{Assel:2010wj}, 
fractionally charged particles are always present in the massless spectrum, with very few
exceptions, and usually not all of them are coupled to other non-abelian interactions to confine them.
The fact that examples can be found where they are absent does not change the fact that string theory 
seems to predict them.

Our results demonstrate the existence of hundreds of novel heterotic ``mini-landscapes" where the 
standard model might be located, as shown in Table \ref{HWLSummary}. Further study of the 
feasibility of this class require methods beyond RCFT. In particular, a geometric understanding of this
class would be most welcome.

Presumably all this is just the tip of a huge iceberg. Along the same lines, one could still study 
weight lifting for exceptional $N=2$ MIPFs. Indeed,
combining   these two ideas is completely straightforward, but we have not done this yet. 
One can also construct internal sectors of heterotic strings partly out of free
fermions and partly out of lifted Gepner models. This also yields new examples of 
three family models \cite{Netjes}. Another possibility
is to use double lifts, as explained in \cite{GatoRivera:2009yt} 
(so far a partial investigation did not yield examples with three families).

But the most important reason to think that there may be a large number of other 
possibilities is that the list of lifted Gepner models
is not known to be complete, and presumably only contains the simplest 
possibilities, those that can be obtained by fairly
straightforward manipulations of affine Lie algebras.

\vskip 2.truecm
\noindent
{\bf Acknowledgements:}
\vskip .2in
\noindent
This work has been partially 
supported by funding of the Spanish Ministerio de Ciencia e Innovaci\'on, Research Project
FPA2008-02968, and by the Project CONSOLIDER-INGENIO 2010, Programme CPAN
(CSD2007-00042). The work of A.N.S. has been performed as part of the program
FP 57 of Dutch Foundation for Fundamental Research of Matter (FOM). 

\bibliography{REFS}
\bibliographystyle{lennaert}

\end{document}